\documentclass[
 reprint,
longbibliography ,
 aps,
 twoside,
 superscriptaddress, 
 bibnotes,
]{revtex4-1}

\usepackage{physics}
\usepackage{dsfont} 
\usepackage{color,newfloat}
\usepackage{graphicx}
\usepackage{dcolumn}
\usepackage{bm}
\usepackage{amsmath,amssymb}
\usepackage{soul}
\usepackage{upgreek}
\usepackage[dvipsnames]{xcolor}

\usepackage{float}
\usepackage{chngcntr} 

\usepackage{array}  
\usepackage{hhline}
\usepackage{multirow}
\usepackage{afterpage}
\usepackage[normalem]{ulem}

\usepackage{xspace}
\usepackage{makecell}
\usepackage{array}
\usepackage{float}

\usepackage{natbib}
\usepackage{notoccite}

\newcolumntype{P}[1]{>{\centering\arraybackslash}p{#1}}

\begin{document}

\title{
%
%
%
%
%
%
%
%
Error protected qubits in a silicon photonic chip
}

\date{\today}

\author{Caterina Vigliar}
\affiliation{Quantum Engineering Technology Labs, H. H. Wills Physics Laboratory and Department of Electrical and Electronic Engineering, University of Bristol, BS8 1FD, UK.}

\author{Stefano Paesani}
\affiliation{Quantum Engineering Technology Labs, H. H. Wills Physics Laboratory and Department of Electrical and Electronic Engineering, University of Bristol, BS8 1FD, UK.}

\author{Yunhong Ding}
\email{yudin@fotonik.dtu.dk}
\affiliation{Department of Photonics Engineering, Technical University of Denmark, 2800 Kgs. Lyngby, Denmark.}
\affiliation{Center for Silicon Photonics for Optical Communication (SPOC), Technical University of Denmark, 2800 Kgs. Lyngby, Denmark.}

\author{Jeremy C. Adcock}
\affiliation{Quantum Engineering Technology Labs, H. H. Wills Physics Laboratory and Department of Electrical and Electronic Engineering, University of Bristol, BS8 1FD, UK.}

\author{Jianwei Wang}
\email{jianwei.wang@pku.edu.cn}
\affiliation{State Key Laboratory for Mesoscopic Physics, School of Physics, Peking University, Beijing, China.}
\affiliation{Frontiers Science Center for Nano-optoelectronics \& Collaborative Innovation Center of Quantum Matter, Peking University, Bejing, China.}

\author{Sam Morley-Short}
\affiliation{Quantum Engineering Technology Labs, H. H. Wills Physics Laboratory and Department of Electrical and Electronic Engineering, University of Bristol, BS8 1FD, UK.}

\author{Davide Bacco}
\affiliation{Department of Photonics Engineering, Technical University of Denmark, 2800 Kgs. Lyngby, Denmark.}
\affiliation{Center for Silicon Photonics for Optical Communication (SPOC), Technical University of Denmark, 2800 Kgs. Lyngby, Denmark.}

\author{Leif K. Oxenl{\o}we}
\affiliation{Department of Photonics Engineering, Technical University of Denmark, 2800 Kgs. Lyngby, Denmark.}
\affiliation{Center for Silicon Photonics for Optical Communication (SPOC), Technical University of Denmark, 2800 Kgs. Lyngby, Denmark.}

\author{Mark G. Thompson}
\affiliation{Quantum Engineering Technology Labs, H. H. Wills Physics Laboratory and Department of Electrical and Electronic Engineering, University of Bristol, BS8 1FD, UK.}

\author{John G. Rarity}
\affiliation{Quantum Engineering Technology Labs, H. H. Wills Physics Laboratory and Department of Electrical and Electronic Engineering, University of Bristol, BS8 1FD, UK.}

\author{Anthony Laing}
\email{anthony.laing@bristol.ac.uk}
\affiliation{Quantum Engineering Technology Labs, H. H. Wills Physics Laboratory and Department of Electrical and Electronic Engineering, University of Bristol, BS8 1FD, UK.}

\begin{abstract}
  \noindent
  General purpose quantum computers
  can, in principle,
  entangle a number of noisy
  physical qubits
  to realise composite
  qubits protected against errors.
  %
  Architectures for measurement-based quantum computing
  intrinsically support error-protected qubits
  and
  are the most viable approach for constructing an
  all-photonic quantum computer.
  %
  %
  %
  %
  %
  %
  %
  %
  Here
  we propose and demonstrate
  an integrated silicon photonic architecture
  that both
  entangles multiple photons,
  and encodes multiple physical qubits on individual photons,
  to produce error-protected qubits.
  %
  %
  %
  We realise reconfigurable graph states
  to compare several schemes with and without
  error-correction encodings
  and implement a range of quantum information processing tasks.
  %
  We observe a success rate increase
  from 62.5\% to 95.8\%
  when running a phase estimation algorithm
  without and with error protection, respectively.
  Finally,
  we realise hypergraph states,
  which
  are a generalised class of resource states
  that offer protection against correlated errors.
  Our results
  show how quantum error-correction encodings
  can be implemented with 
  resource-efficient photonic architectures
  to improve the performance of quantum algorithms.
\end{abstract}

\maketitle

\begin{figure*}[ht]
  \centering
  \includegraphics[trim=0 0 0 10, width=0.91 \textwidth]{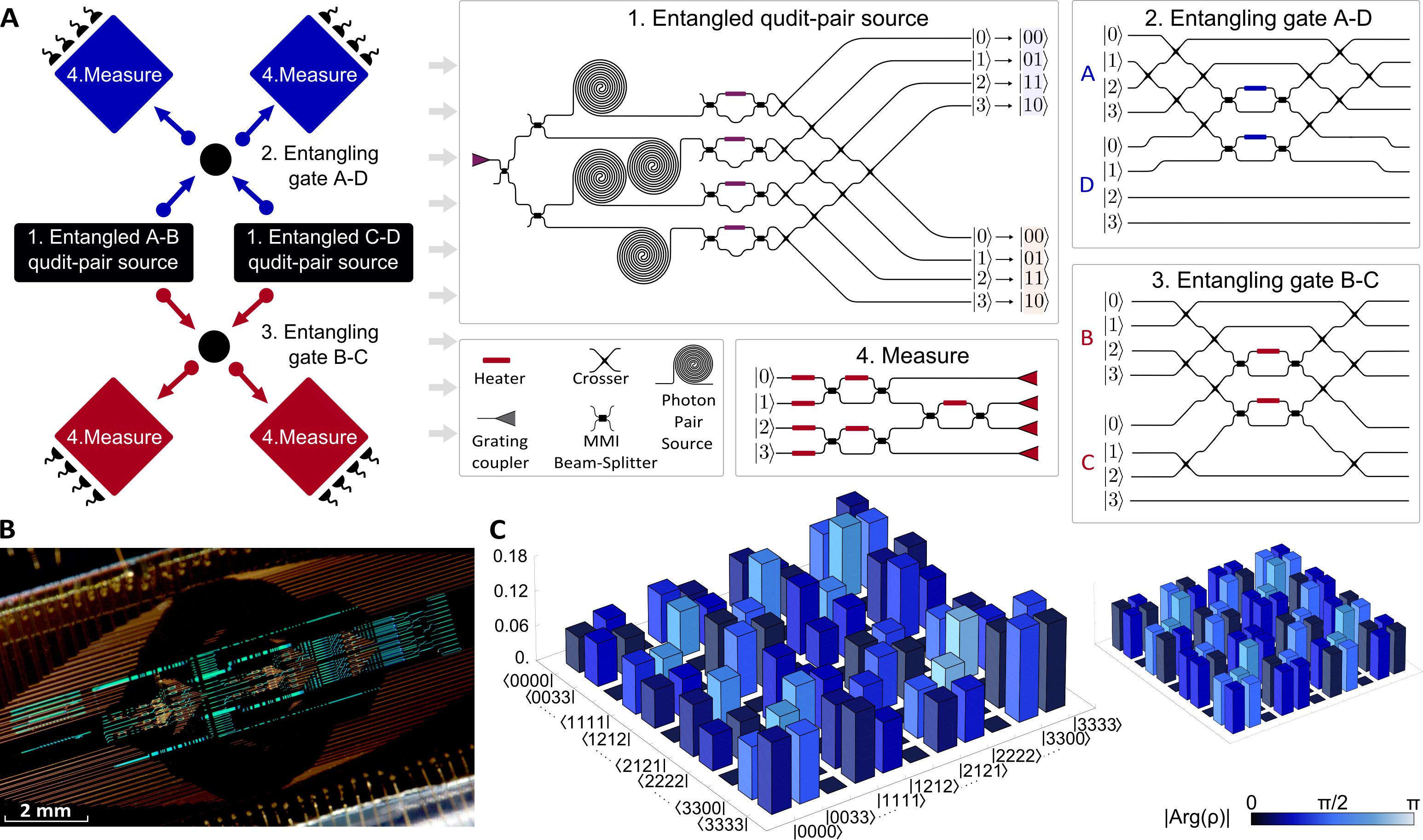}
  \caption{
  \textbf{Device description and performance.}
  \textbf{a.} Schematics of the silicon chip modules. 
  Two entangled qudit-pair modules (black rectangles),
  each consisting of four photon-pair sources,
  generate two pairs of maximally entangled qudits
  (A-B and C-D).
  %
  Each qudit, encoded in either a signal (red)
  or an idler (blue) photon,
  is mapped to a two-qubit system.
   Photons A-D and B-C are entangled 
   with fusion-type gates (black circles)
   to produce entangled states of 8 qubits among 4 parties.
  Measurement modules (blue and red squares)
  perform arbitrary projective measurements on each photon.
  \textbf{b.} Photograph of the silicon photonic chip. 
  \textbf{c.} State characterisation of a trial four-photon four-dimensional entangled state: $(\ket{0000} + e^{-i\pi/4}\ket{0033} + e^{i\pi/2}\ket{1111} + \ket{1212} + e^{i\pi/4}\ket{2121} + e^{-i\pi/4}\ket{2222} + e^{i\pi/2}\ket{3300} + e^{i\pi/4}\ket{3333})/2^{3/2}$. The non-zero elements of the density matrix (left) are measured to obtain  fidelity of $0.72 \pm 0.04$
  with the theoretical target state (right).
  %
  %
  %
  Column heights represent the absolute value of the non-zero entries of the complex matrix, while phases are colour coded. Error bars are obtained from Monte Carlo simulations assuming a Poissonian distribution of the measured counts.
  }
\label{FigSetUp}
\end{figure*}

\noindent
While rudimentary quantum computers
can now solve abstract tasks
that are intractable to classical computers~\cite{arute2019GoogleSupremacy,Bernien2017,preskill2018},
a general purpose quantum computer
will require error-correcting schemes to protect information as it is processed in a useful quantum algorithm \cite{gottesman1997stabilizer,aharonov1999faultTol}.
%
%
%
%
Measurement-based quantum computing (MBQC)
uses entangled states of physical qubits,
namely graph states,
to run an algorithm by
measuring sequences of qubits
and propagating information
between different physical layers of the state~\cite{Raussendorf2001, Raussendorf2003, walther2005experimental, vallone2008activeOWQC, wang201816Supercondicting}. 
As graph states are both the resource for MBQC
and have a direct mapping to quantum error correcting codes for 
error-protected qubits,
MBQC intrinsically facilitates a fault-tolerant architecture~\cite{EncodedMBQCZwerger, raussendorf2007topological, schlingemann2001ECCAndGraphs, MBErrCorrTrappIonsLanyon, yao2012experimental, bell2014experimental}.
%
%

Integrated photonics
is an appealing platform for 
architectures that build
large graph states
for fault-tolerant MBQC
\cite{Varnava2008,Segovia2015,RudolphOptimistic}.
%
%
%
Thousands of optical components can be densely integrated onto a single silicon chip~\cite{16DPapero, Harris2017, AGGJeremy, llewellyn2019chip}, to realise the complex photonic circuitry
required to generate and control entangled states of many photons.
%
%
Encoding multiple qubits onto individual photons,
within a multi-photon state,
provides a natural resource-saving approach
\cite{16DPapero}.
%

%
Here we develop an architecture
that combines on-chip generation of multiple
pairs of entangled photons and
encoding of multiple qubits
on individual photons
via the creation of $d$-level systems---qudits---to realise programmable graph states
and 
error-protected qubits. 
%
%
%
Small clusters of qubits
are encoded in single photons;
multiple photons are then fused together
with probabilistic entangling gates.
Our architecture is modular,
with the production, processing, and measurement
of the graph states performed by
different units of the same chip.

We experimentally implement this architecture on a silicon photonic chip to realise eight-qubit reconfigurable graph states encoded in four photons.
Each photonic module
embeds tens of optical components,
totalling more than 220 components,
including 8 photon-pair sources
and programmable circuitry with
48 phase shifters.
We experimentally investigate
different classes of graph states and
demonstrate error-protected computations
and teleportation schemes.
In our device, the error-protection of measurement-based operations is performed via a probabilistic correction of computational errors affecting physical nodes in the graph state.
We run a simple version of the phase estimation algorithm and demonstrate the improvement in performance
that results from implementing 
error-correction encodings.
Finally, we experimentally demonstrate the generation and processing of hypergraph states---generalised graph state resources that can represent a novel approach to the MBQC paradigm.\\

\noindent
\textbf{Architecture and device}\\
\noindent
Our silicon photonic chip, shown in Fig.~\href{\ref{FigSetUp}}{\ref{FigSetUp}b},
can be decomposed into three different modules, as shown in Fig.~\href{\ref{FigSetUp}}{\ref{FigSetUp}a}. 
First, eight coherently-pumped spiral waveguides ($1.5$ cm long) generate two maximally-entangled pairs of photons,
which are spatially separated with integrated filters
based on asymmetric Mach-Zehnder interferometers
and waveguide crossers \cite{16DPapero}.
Each photon encodes a pair of qubits using the
spatial mode mapping
$\{\ket{0} \rightarrow \ket{00},
\ket{1} \rightarrow \ket{01},
\ket{2} \rightarrow \ket{10},
\ket{3} \rightarrow \ket{11}\}$,
as shown in Fig.~\href{\ref{FigSetUp}}{\ref{FigSetUp}a},
to realise an eight-qubit system. 
Next, programmable entangling gate modules (shown in Fig.~\href{\ref{FigSetUp}}{\ref{FigSetUp}a}) are used to
realise graph states of these
eight-qubits
implementing fusion operations 
between the different photon pairs~\cite{zhang2008FusionGate, AGGJeremy}.
%
%
Finally, a triangular structure of MZIs performs
arbitrary projective measurements.
%
Photons are fibre coupled off-chip using low-loss ($<1$~dB) grating couplers~\cite{ding2014Gratings}, and routed to high-efficiency superconducting nanowire single photon detectors (SNSPDs) (see Supplementary Sections~\ref{SectionAppendixExpSetup} and~\ref{SectionAppendixDevicePerformance}
for further experimental details).

\begin{figure}[b!]
  \centering
  \includegraphics[
  trim=0 0 0 15,
  width=0.47\textwidth]{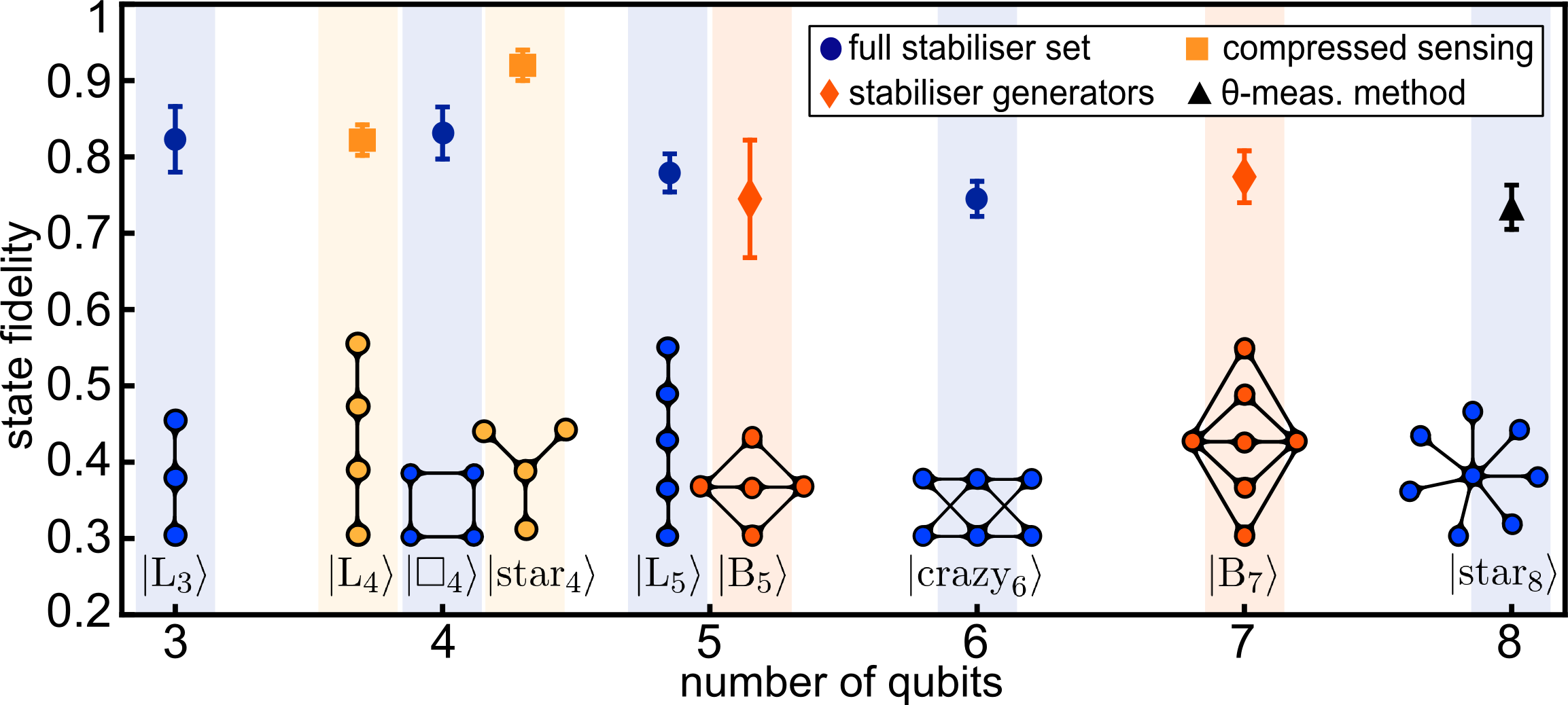}
  \caption{
\textbf{Graph state fidelities.} 
Estimates by stabiliser methods are shown as blue dots,
while orange diamonds show the mean of only stabilizer generator measurements. 
Fidelity estimations by compressed sensing tomographies
are indicated by yellow squares (two-photon data). 
A fidelity of $0.73 \pm 0.03$ for the eight-qubit star graph, obtained via $\theta$-measurements methods, is shown as a black triangle. 
All error bars are obtained assuming a Poissonian distribution of the measured counts.
} 
  \label{FigGraphStates}
\end{figure}
\begin{figure*}[ht!]
  \centering
  \includegraphics[trim=0 0 0 10, width=1 \textwidth]{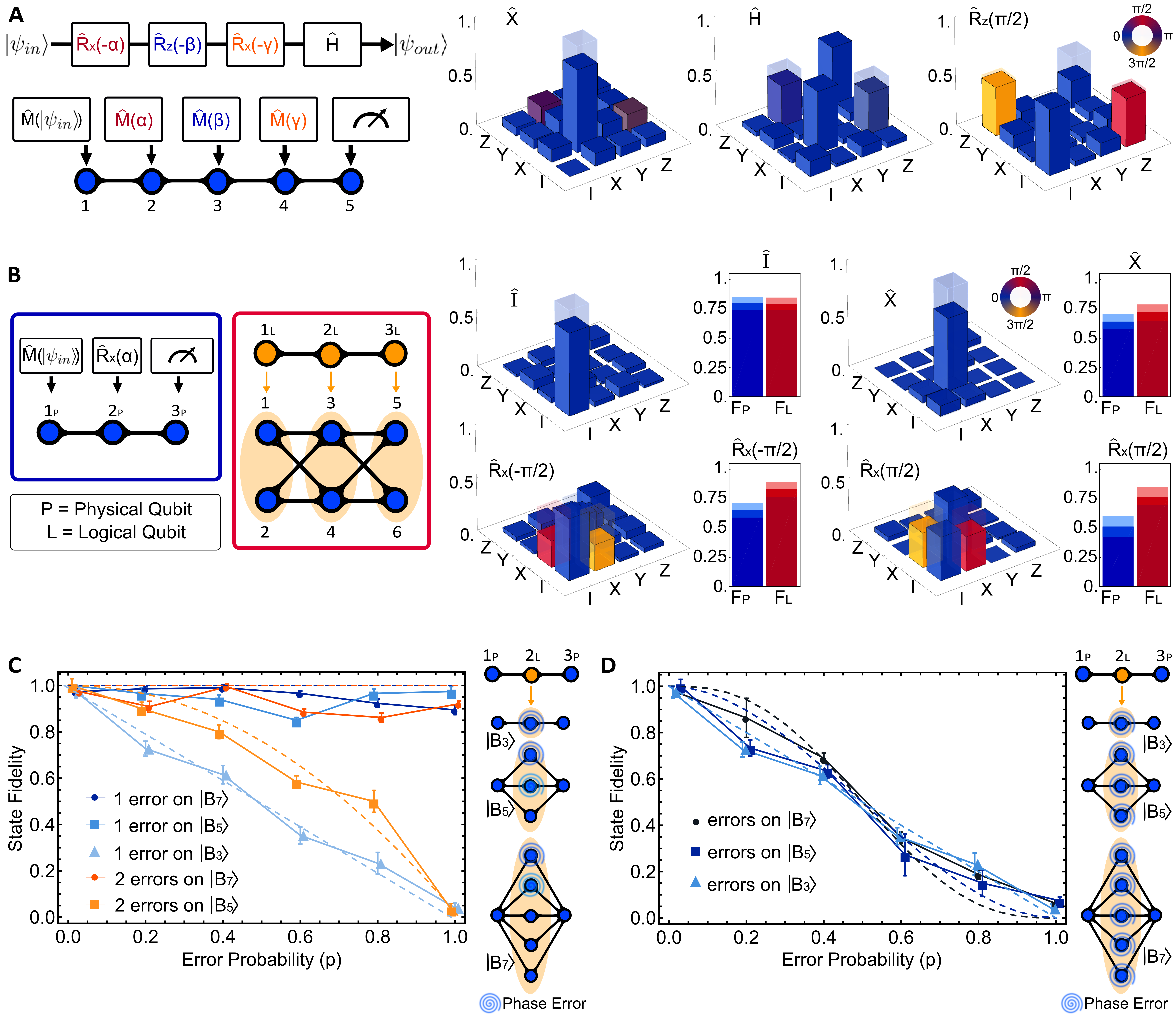}
  \caption{
  \textbf{MBQC operations with physical and logical graph states.}
  \textbf{a.} Single qubit operations.
  The schematics show
  preparation, unitary gate, and read-out
  using a five-qubit linear cluster,
  together with a circuit model representation.
  %
  %
  Process tomography data for the
  $\hat{X}$, Hadamard and $\hat{R}_Z({\pi\over2})$ gates
  give respective fidelities of
  $0.79 \pm 0.06$, $0.98 \pm 0.08$ and $0.92 \pm 0.06$,
  with the ideal cases shown in transparency. 
  Column heights represent the absolute values of the complex matrix entries and phases are colour coded.
  \textbf{b.} MBQC using physical and logical qubits.
   Schematics show
   a single qubit process
   using MBQC with
   three physical qubits (blue frame)
   and three logical qubits (red frame).   
  Each logical qubit is encoded on two physical qubits
  to protect against phase-flip errors.
  %
  Matrices for single qubit processes implemented
  with logical-qubit MBQC are shown for 4 operations;
  their fidelities (red) are compared against
  the same operations implemented with physical qubits (blue), where lighter-coloured regions represent error bars. 
  \textbf{c.} Results for measurement-based error-corrected
  state teleportation.
  Branched graph states comprising one-, three- and five-qubit repetition codes, are labelled $\ket{\mathrm{B}_3}$, $\ket{\mathrm{B}_5}$ and $\ket{\mathrm{B}_7}$ respectively. 
  Fidelities are reported against
  increasing probability of a phase error occurring 
  on one and two physical qubits
  of the code respectively, realised by detuning the measurement basis.
  \textbf{d.} (As for \textbf{c}) Dephasing errors affecting all physical qubits in the code. Ideal cases are shown as dashed lines. 
  Fidelity improvements when increasing the repetition code size are experimentally observed for error rates below the noise threshold $p=0.5$ (up to $14\%$ for $p=0.2$), while no benefits are observed above this threshold.  
  All error bars represent the standard error of the mean, obtained from Monte Carlo simulations assuming a Poissonian distribution of the measured counts.
  }
  \label{FigErrorCorrection}
\end{figure*}
\begin{figure*}[ht!]
  \centering
  \includegraphics[trim=0 0 0 10, width=1. \textwidth]{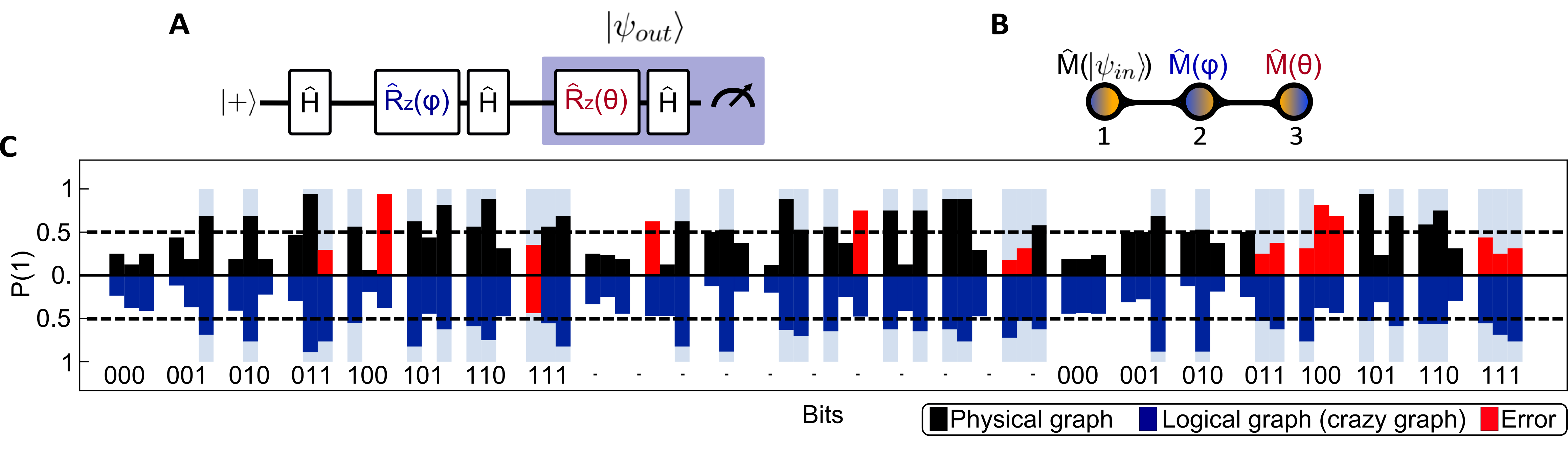}
  \caption{
  \textbf{Experimental results for MBQC phase estimation algorithm with physical and logical graphs.}
  %
  %
  %
  %
  %
  %
  \textbf{a.} Circuit representation of the entanglement-free version of PEA, which (\textbf{b}) can be mapped into a MBQC implementation using a three-qubit line graph of physical or logical qubits. 
  \textbf{c.} Results for phase estimation of three-bit phases using the physical (top half, black) and logical (bottom half, blue) line graph states. The probability P(1) of measuring the rightmost logical qubit in the output 1 is plotted, from which we infer the phase bit 0 (P$(1)<0.5$) or 1 (P$(1)>0.5$).
  Failures in the phase estimation are highlighted as red bars, while light blue bars show the correct bit values.
  } 
  \label{FigMBAlgos}
\end{figure*}
%
%
%

Different graph states can be realised from different four-photon four-dimensional states, that are obtained by pumping different sources and by programming both the entangling gates and the single-photon two-qubit measurement stages.
The characterisation of one of the initial
four-photon four-dimensional 
resources,
using partial state tomography~\cite{malik2016multiPmultiD},
is shown in
Fig.~\href{\ref{FigSetUp}}{\ref{FigSetUp}c},
for which we measure
a fidelity of $0.72 \pm 0.04$
against the ideal state 
(see Supplementary Section~\ref{SectionAppendix4P4DFidelity}
for more details).
A larger variety of graph state topologies is available by encoding multiple qubits onto single photons than is available by encoding a single qubit per photon and using post-selected linear optical operations \cite{adcock2018postselection,gu2019}.
In fact, local operations on individual photons, each encoding four-dimensional qudits,  allow non-local operations between the two qubits encoded in that photon.
Such non-local operations are deterministic and have significantly higher fidelities than gates between different photons (due to the absence of distinguishability limitations in multi-photon interference).
These properties are advantageous for encoding physical qubits that make up a single logical node within an individual photon. When using codes with constant numbers of qubits, e.g. five-qubit codes to correct arbitrary local unitary errors~\cite{EncodedMBQCZwerger}, the overhead of such an encoding is a constant number of additional modes per photon, which is practical for current integrated quantum photonics technologies~\cite{16DPapero}.
Overall, our chip generates 21 graph state equivalence classes with up to 8 qubits (more than 120 different graphs). 
This is in contrast to the 4 classes (8 different graphs) with up to 4 qubits that are accessible with two-dimensional photonic architectures using post-selection and the same number of processed photons
(for more details, see Supplementary Section~\ref{SectionAppendixGraphStates}). 

%
%
%
%

Figure~\ref{FigGraphStates} reports fidelities
for a variety of graph states
with up to eight qubits.
These include the eight-qubit star graph ($\ket{\mathrm{star}_8}$), 
the ``double-branched" graph states
of seven ($\ket{\text{B}_7}$)
and five ($\ket{\text{B}_5}$) qubits,
the six-qubit ``crazy graph''
($\ket{\text{crazy}_6}$) \cite{RudolphOptimistic},
the five-, four- and three-qubit
linear states ($\ket{\text{L}_5}$, $\ket{\text{L}_4}$
and $\ket{\text{L}_3}$, respectively),
and the four-qubit star ($\ket{\mathrm{star}_4}$)
and box ($\ket{\Box_4}$) states.
States shown in Fig.~\ref{FigGraphStates} are all derived from ($\ket{\mathrm{star}_8}$),
via local rotations and two-qubit (intra-qudit) operations.
Our graph state verification methods (stabilizer measurements \cite{guhne2009entanglement},
compressed sensing tomography \cite{gross2010CompressedSensing}
and $\theta$-measurements \cite{sackett2000GHZVerif})
and the device configurations
used to generate these states
are discussed in
Supplementary Section~\ref{SectionAppendixGraphStates}.\\

\noindent
\textbf{Physical and logical MBQC}\\
\noindent
Different classes of graph states correspond to different computational tasks in the MBQC paradigm \cite{Raussendorf2003, hein2004multiparty}.
The ability to reprogram graph states enables us to reprogram the type of computation.
As shown in
Fig.~\href{\ref{FigErrorCorrection}}{\ref{FigErrorCorrection}a},
we tested universal single qubit gate operations
in MBQC
with arbitrary state initialisation and measurement,
using
a five-qubit one-dimensional lattice.
Measurements on the first and last physical qubits
perform the state preparation and readout,
while measurements on the three central qubits
implement Euler decompositions of arbitrary single qubit gates \cite{Raussendorf2003}.
The ability to initialise and read-out in arbitrary bases allows us to characterise this MBQC channel via full process tomography \cite{NielsenChuang2002}
(for further details
see Supplementary Section~\ref{SectionAppendixEncodingMBW} and~\ref{SectionAppendixErrorCorrMBQC}).
Fig.~\ref{FigErrorCorrection}a
shows single-qubit MBQC process tomography
for different gates:
a Hadamard gate,
a rotation $\hat{R}_Z({\pi\over2})$,
and an $\hat{X}$ gate.
The measured fidelities for these gates are
$0.98 \pm 0.08$, $0.92 \pm 0.06$, and $0.79 \pm 0.06$, respectively.
Error bars represent the standard error of the mean, obtained from Monte Carlo simulations assuming a Poissonian distribution of the measured counts.

%
%

Logical states and operations can be protected in MBQC by replacing each physical qubit of the lattice with a stabiliser error-correcting code. 
This results in a new graph definition
in which each vertex is a logical qubit,
encoded in multiple physical qubits;
%
the overall state of physical qubits
can still be represented as a graph state \cite{EncodedMBQCZwerger}. 
For example,
so called \emph{crazy graph} structures
can equip linear graph states
with error-correction,
where a repetition code
associated to each column of the two-dimensional physical lattice replaces a single  physical vertex, resulting in a linear logical graph~\cite{RudolphOptimistic}.
%
Such states are suitable for photonic fault-tolerant architectures as they can correct both computational errors and high levels of qubit loss~\cite{morleyShort2019LossTol}.
An example of crazy graph encoding
is represented in Fig.~\href{FigErrorCorrection}{\ref{FigErrorCorrection}b}
for the case of a three-qubit logical linear graph state
encoded in the six-qubit crazy graph physical state. 
In this case, each logical qubit of the linear graph is encoded in a two-qubit repetition code in the $\hat{X}$ basis to detect phase-flip errors~\cite{NielsenChuang2002}:
$ \ket{+}_{i_\mathrm{L}}=\ket{++}_{2i-1,2i}$, 
$\ket{-}_{i_\mathrm{L}}=\ket{--}_{2i-1,2i}$, 
%
where $i\in \{1,2,3\}$ indicates the column of the crazy graph.
Cases such as $\ket{+-}_{ij}$ are detected as errors and filtered out.
Note that local measurements in the logical computational basis correspond to non-local measurements of physical qubits encoded in the same qudit.
Using 
the crazy graph 
state we perform process tomography of single-qubit $\hat{R}_X$ rotations in MBQC on a logical lattice. 
Error protection is performed throughout the whole process of the computation: state encoding, processing, and read-out.
Tomography matrices measured for different processes
are reported in Fig.~\href{FigErrorCorrection}{\ref{FigErrorCorrection}b} for both uncorrected and error-corrected cases.
Crucially, for each gate analysed
the fidelity with error protection exceeds that without
error protection.
See Supplementary Sections~\ref{SectionAppendixErrorCorrMBQC} and~\ref{SectionAppendixPhysicalErrors} for further details on the tomography procedures and on how physical noises map to computational errors in the different encodings used.

%

%
More resilient error correcting codes can be realised by using larger repetition codes \cite{NielsenChuang2002, schindler2011expBlattRepCode, MBErrCorrTrappIonsLanyon}.
We implement such codes using the branched states $\ket{\mathrm{B}_5}$ and $\ket{\mathrm{B}_7}$, equivalent to a three-qubit linear graph where the middle vertex is protected against phase flip errors via three-qubit and five-qubit repetition codes, respectively.
We test the performance of these codes against the uncorrected case ($\ket{\mathrm{B}_3}$) by implementing state teleportation of the state $\ket{+i} = \frac{1}{\sqrt{2}}(\ket{0}+i\ket{1})$ between the two outer (unprotected) qubits of the branched graphs. 
Fig.~\href{FigErrorCorrection}{\ref{FigErrorCorrection}c} shows results for the behaviour of the graphs under different fault scenarios.
In particular, we test the cases where errors affect one and two physical qubits of the intermediate layer with error probabilities $0\geq p \geq1$.
A more practical scenario where all qubits of the intermediate layer are equally affected with error probability $p$ is reported in Fig.~\href{FigErrorCorrection}{\ref{FigErrorCorrection}d}.
In all cases,
higher redundancies result in an enhanced fidelity of the teleported state, indicating higher tolerance to computational
errors.
%
%
In the  Supplementary Section~\ref{SectionAppendixSimLossTol} we report additional data on how these graph structures can also tolerate photon loss, even in the case where multiple qubits are encoded on each photon. 
\\ 

\noindent
\textbf{Quantum algorithms with 
error-protected
qubits}\\
\noindent
We used our architecture and device to implement the measurement-based version of the phase-estimation algorithm (PEA), which is a core algorithm in quantum information processing
\cite{OMalley, higgins2007,MBPhaseEst2017Friis, Paesani2017}.
\begin{figure}[t!]
  \centering
  \includegraphics[trim=0 0 0 10, width=0.9\linewidth ]{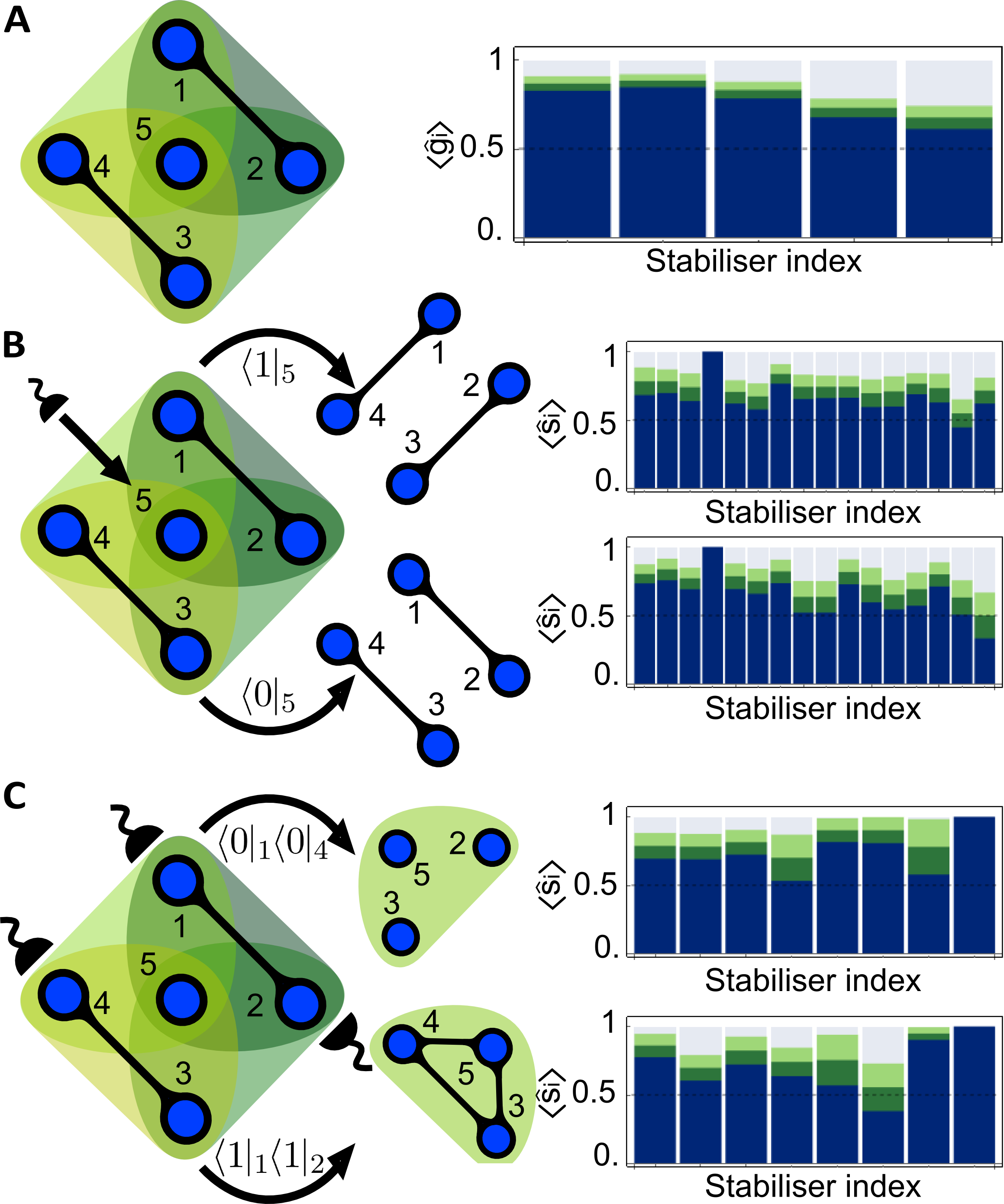}
  \caption{
  \textbf{Results for MBQC with hypergraphs.}
  \textbf{a.} Clover hypergraph and measured stabiliser generators, with $0.81 \pm 0.02$ mean expectation values.
  \textbf{b.} Effects of measuring the central qubit of the clover state in the $\hat{Z}$ basis. 
  A measurement outcome 1 induces a CZ operation within vertices sharing a hyperedge, resulting in two Bell pairs (1-4) (3-2), while outcome 0 induces no operations, leaving the two Bell pairs (1-2) (3-4).
  Stabiliser expectation values for the resulting states are shown on the right, providing fidelities $0.74 \pm 0.03$ and $0.74 \pm 0.03$, respectively. 
  \textbf{c.} Effects of measuring two outer qubits of the clover state in the $\hat{Z}$ basis, leading to the generation of the Toffoli and fully-connected Toffoli hypergraphs. 
  Stabiliser measurements provide fidelity estimates of $0.83\pm 0.05$ and $0.80 \pm 0.05$, respectively. 
  All error bars are obtained assuming a Poissonian distribution of the measured counts.
  }  
  \label{FigHyperGraphs}
\end{figure}
%
%
%
%
We experimentally investigated a rudimentary form of error-correction on a simple version of PEA 
that does not require entanglement
with ancillary qubits \cite{higgins2007}.
The goal of PEA can be stated as reconstructing a phase $2\pi\phi_0 $ (with $\phi_0\in[0,1)$) acting on an input quantum state.
An iterative version of PEA does so by iteratively inferring---in reverse order---the bits in the binary expansion of $\phi_0 $ using the logical circuit in Fig.~\href{FigMBAlgos}{\ref{FigMBAlgos}a}, where the rotation angles in the gates $\varphi=2\pi M \phi_0$ and $\theta$ are adaptively chosen during the inference process \cite{dobvsivcek2007IterativeQPEA, kitaev1995IterativeQPEA}.

In the MBQC implementation, the phases $\varphi$ and $\theta$ are realised by rotating the measurement bases on the physical qubits in the graph, as shown in Fig.~\href{FigMBAlgos}{\ref{FigMBAlgos}b}.
The protocol can be performed on a three-qubit line graph, which can be realised with either three ($\ket{\text{L}_3}$) or six physical qubits ($\ket{\text{crazy}_6}$),
as described in Supplementary Section~\ref{AppendixMBQCPhaseEst}.
The probabilities of measuring the rightmost qubit in the output 0 or 1, which infers the phase bit 0 or 1, are given by $p=\cos^2[\pi (M \phi_0 - \theta)]$ or $1-p$, respectively.
%
%
Experimental results for error-protected measurement-based PEA, applied to a range of phases with three-bit precision, are shown in Fig.~\href{FigMBAlgos}{\ref{FigMBAlgos}c}, together with results for physical qubits for comparison.
The inference process uses majority voting over a statistical set of 17 samples~\cite{Paesani2017}.
%
%
%
%
%
While in PEA with physical qubits experimental noise results in observing an incorrect phase reconstruction in 9 over the 24 cases tested (62.5\% success rate), using the logical encoding we observe only 1 incorrect phase reconstruction (95.8\% success rate). The statistical confidence associated to observing an improvement in the logical case compared to the physical one is 98\% (see Supplementary Section~\ref{SectionAppendixDataAnalysis} for details on the data analysis).   
Note that, in both physical and 
error-protected
implementations, the experimental resources used are the same (number of photons and number and performance of gates undergone by each photon): the observed improvement is purely provided by a better way of encoding the quantum information in an error-protection code.\\

\noindent\textbf{Hypergraphs}\\
 Generalisations of MBQC have recently been proposed based on the more general class of \emph{hypergraph} states \cite{Qu2013,Rossi2013,guhne2014EntPropHyper}.
Hypergraph resources can provide advantages in measurement-based protocols, e.g. resource savings and parallelisation~\cite{miller2016OnlyPauliMeasOnHyperGr, Gachechiladze2016MetrologyHyper,gachechiladze2019circuitDepthHYPER}, protection against correlated errors~\cite{lyons2017local} and improved noise robustness in quantum metrology applications \cite{Gachechiladze2016MetrologyHyper}.
The hypergraph state formalism allows for the presence of higher order correlations, which are visually represented as loops surrounding vertices, called hyperedges \cite{Qu2013,Rossi2013,guhne2014EntPropHyper, gachechiladze2017graphicalLCHYPER}. 
More specifically, a loop around $k$ qubits labelled $\{i_1, i_2, ...i_k\}$ describes a generalised $k$-qubit controlled-phase (CZ) operation, C$^{k}$Z, where C$^{k}$Z $= \hat{I}-2\ket{1...1}{}_{i_1 ... i_k}\bra{1...1}$.
The number of vertices contained in a hyperedge defines its cardinality.
As for graph states,
a (non-local) stabiliser formalism can be introduced for hypergraphs \cite{morimae2017verificationHYPER},
which enables state characterisation with the same techniques.

Our device can be programmed to generate a variety of hypergraph states of up to five-qubits.
We test the ``clover" five-qubit hypergraph state,
represented in Fig.~\href{FigHyperGraphs}{\ref{FigHyperGraphs}a}, which contains two edges of cardinality two and four hyperedges of cardinality three
(see Supplementary Section~\ref{SectionAppendixGraphStates} for details).  
The measured mean of the expectation values of the state's stabiliser generators is $0.81 \pm 0.02$, giving an indication of its quality.

Basic working principles of MBQC using hypergraphs can be demonstrated using the clover hypergraph.
In particular, we experimentally investigate the action of $\hat{Z}$ measurements.
While in standard MBQC on graph states a $\hat{Z}$ measurement removes a physical qubit from the graph, up to probabilistic local Pauli gates on the neighbouring vertices, in hypergraphs the resulting probabilistic gates are non-local operations between the vertices sharing a hyperedge with the removed qubit \cite{guhne2014EntPropHyper, gachechiladze2019circuitDepthHYPER}. 
This effect is an essential feature arising from high-order non-local correlations in hypergraphs.
For hyperedges of cardinality three, e.g. those present in the clover hypergraph, an outcome 1 in a $\hat{Z}$ measurement induces a CZ operation between the two remaining neighbours, while it induces an identity for outcome 0.

This working principle is demonstrated in the results of Fig.~\href{FigHyperGraphs}{\ref{FigHyperGraphs}b-c}, where the states resulting from different outcomes of $\hat{Z}$ measurements on the clover state are characterised.
Measuring the central qubit of the clover leads to bi-separable states of two Bell-pairs, with the connectivity given by the measurement outcome.
Experimentally, these have an identical state fidelity of $0.74 \pm 0.03$.
A three-qubit ``Toffoli" hypergraph with one single hyperedge, or a ``fully-connected Toffoli'' hypergraph, with one three-qubit hyperedge and three regular edges, are obtained by measuring two outer vertices.
These states have fidelities of $0.83 \pm 0.05$ and $0.80 \pm 0.05$, respectively.\\

\noindent
\textbf{Outlook}\\
%
Although our implementations rely on probabilistic schemes to generate multi-photon states (post-selection and probabilistic entangling gates), the resource savings enabled by our approach could deliver advantages for both noisy intermediate-scale quantum (NISQ) photonic devices in the nearer term, and general-purpose photonic quantum computing architectures over the longer term. 
%
%
The resource savings of high-dimensional encodings open the possibility to develop NISQ systems with up to $\geq 40$ photonic qubits, a scale impractical with previous pre-loss-tolerant photonic approaches
(See Supplementary Section~\ref{SectionAppendixQuditEncScaling}).  
To reach the scale required for universal photonic MBQC, the adoption of fault-tolerant architectures, in particular loss-resilient photonic schemes, is necessary.
Encoding multiple qubits on individual photons allows the encoding of logical qubits that could improve tolerance to logical errors and loss (even if photon loss results in multiple qubits of information being lost)~\cite{Piparo2020}.
%
Crazy graphs are useful architectural building-blocks as they achieve a photon loss-tolerance threshold arbitrarily close to unity with sufficiently large repetition codes~\cite{RudolphOptimistic,morleyShort2019LossTol}. 
In our implementation, 
the first time such states are demonstrated in photonics,
losses are filtered out by post-selection, which precludes a direct exploitation of the loss-resilient properties of the state (see Supplementary Section~\ref{SectionAppendixSimLossTol}).
The bottlenecks of post-selection and probabilistic entangling gates can in principle be addressed with near-deterministic photon generation and gate multiplexing to deliver scalable architectures for photonic quantum computing~\cite{Segovia2015}.
While progress in circuits and sources compatible with such architectures is impressive~\cite{16DPapero,wang2019towards, Paesani2020Sources}, significant challenges remain.  
In particular, the integration of detector arrays, low-loss fast switches that are compatible with on-chip sources, and low-latency control electronics, are key for developing efficient multiplexing of photon sources and gates. 
Recent results in thin-film barium titanate on silicon show great potential for the development of on-chip switches for multiplexing, both demonstrating losses below  $0.5$\,dB at speeds of tens of GHz~\cite{FelixEltes2019integrated}. 
Low-latency feed-forward techniques used for multiplexing, possible for example via the integration or interconnection of CMOS electronics with the photonic platform,  will also enable scaling error-corrected measurement-based quantum photonic operations~\cite{Segovia2015,atabaki2018integrating}.

The results we report provide experimental evidence that 
error protection
can give advantages not only for fault-tolerant machines, but also for near-term applications on pre-fault-tolerant devices.
Another way of understanding our results is that we have demonstrated improved process fidelity by increasing the width of optical circuitry for 
error-protected
qubit operation, while maintaining the same optical depth as that used for physical qubit operation.
Combining recent technological advances in high-dimensional encoding and multiphoton capability with integrated quantum photonics \cite{16DPapero,paesani2019}, our architecture can readily provide reconfigurable photonic graph states with tens of qubits. \\



\bibliography{biblio}

\begin{footnotesize}

\hfill \vspace{-0.1cm}

\noindent
\textbf{Acknowledgements.}
\noindent We thank R. Santagati, A. E. Jones, J. F. Bulmer, R. Shaw, J. F. Tasker, N. Maraviglia, J. W. Silverstone, W. A. Murray, Z. Raissi, C. Gogolin and P. Skrzypczyk for useful discussions and technical assistance.
We acknowledge support from the 
Engineering and Physical Sciences Research Council (EPSRC), European Research Council (ERC), and European Commission (EC) funded grants PICQUE, BBOI, QuChip, QuPIC, QITBOX, {VILLUM FONDEN, QUANPIC (ref. 00025298)}, the Center of Excellence, Denmark SPOC (ref DNRF123) and EraNET cofund initiatives QuantERA within the European Union’s Horizon
2020 research and innovation program grant agreement No.
731473 (project SQUARE).
Fellowship support from EPSRC is acknowledged by A.L. (EP/N003470/1).\\

\noindent\textbf{Data availability.} 
The data that support the plots within this paper and other
findings of this study are available at https://doi.org/10.6084/m9.figshare.11903427.\\

%
%

%
\noindent\textbf{Competing financial interests.} M.T. is involved in developing quantum photonic technologies at PsiQuantum Corporation.\\

\end{footnotesize}

\clearpage
\newpage 

\pagenumbering{arabic}
\onecolumngrid
\setcounter{page}{1}

\renewcommand{\thefigure}{\arabic{figure}}
\renewcommand{\figurename}{Supplementary Figure}
\setcounter{figure}{0} 

\renewcommand{\thetable}{\arabic{table}}
\renewcommand{\tablename}{Supplementary Table}
\setcounter{table}{0}

\counterwithout{equation}{section}
\renewcommand{\theequation}{S\arabic{equation}}
\setcounter{equation}{0}

\renewcommand{\thesection}{\arabic{section}}
\renewcommand{\thesubsection}{\alph{subsection}}

\begin{center}
  {\fontsize{15}{15}\selectfont 
\textbf{Supplementary Information}}  
\end{center}

\section{Device and experimental setup} \label{SectionAppendixExpSetup}

\subsection{Device fabrication} 
The fabrication process begins with a commercial silicon-on-insulator (SOI) wafer with top silicon thickness of 250\,nm and buried oxide layer of 3\,$\upmu$m, and proceeds in the following steps.
First, a 1.6\,$\upmu$m thick layer of SiO$_2$ is deposited on the SOI wafer by plasma-enhanced chemical vapour deposition (PECVD).
Next, an Al mirror is deposited by an electron-beam (e-beam) evaporator, followed by another layer of SiO$_2$ of 1\,$\upmu$m thickness.  
Following that, the wafer is flip-bonded with benzocyclobutene bonding process to another silicon carrier wafer. 
The substrate and buried oxide layers of the original SOI wafer are removed by fast dry-etching and buffered hydrofluoric acid chemical-etching, resulting in a Al-introduced SOI wafer~\cite{ding2014Gratings}.
After that, the silicon photonic circuit is fabricated by standard e-beam lithography (EBL) followed by inductively coupled plasma etching and e-beam resist stripping. 
The grating couplers on the device are fully-etched, meaning only one silicon etch step is required.

Once the waveguides layer is fabricated, 1.3\,$\upmu$m thick SiO$_2$ is deposited by PECVD, followed by a polishing process to planarize the surface with approximately 300\,nm sacrifice, resulting in a final top SiO$_2$-cladding layer of 1\,$\upmu$m. 
Afterwards, the phase shifters are patterned by an EBL process.
This is followed by deposition of a 100\,nm titanium (Ti) layer and liftoff process. 
The conducting wires and electrode pads are obtained by standard ultraviolet lithography followed by Au/Ti deposition and lift-off process.\\

\begin{figure*}[b]
 \centering
 \includegraphics[trim=0 0 0 10, width=0.9 \textwidth]{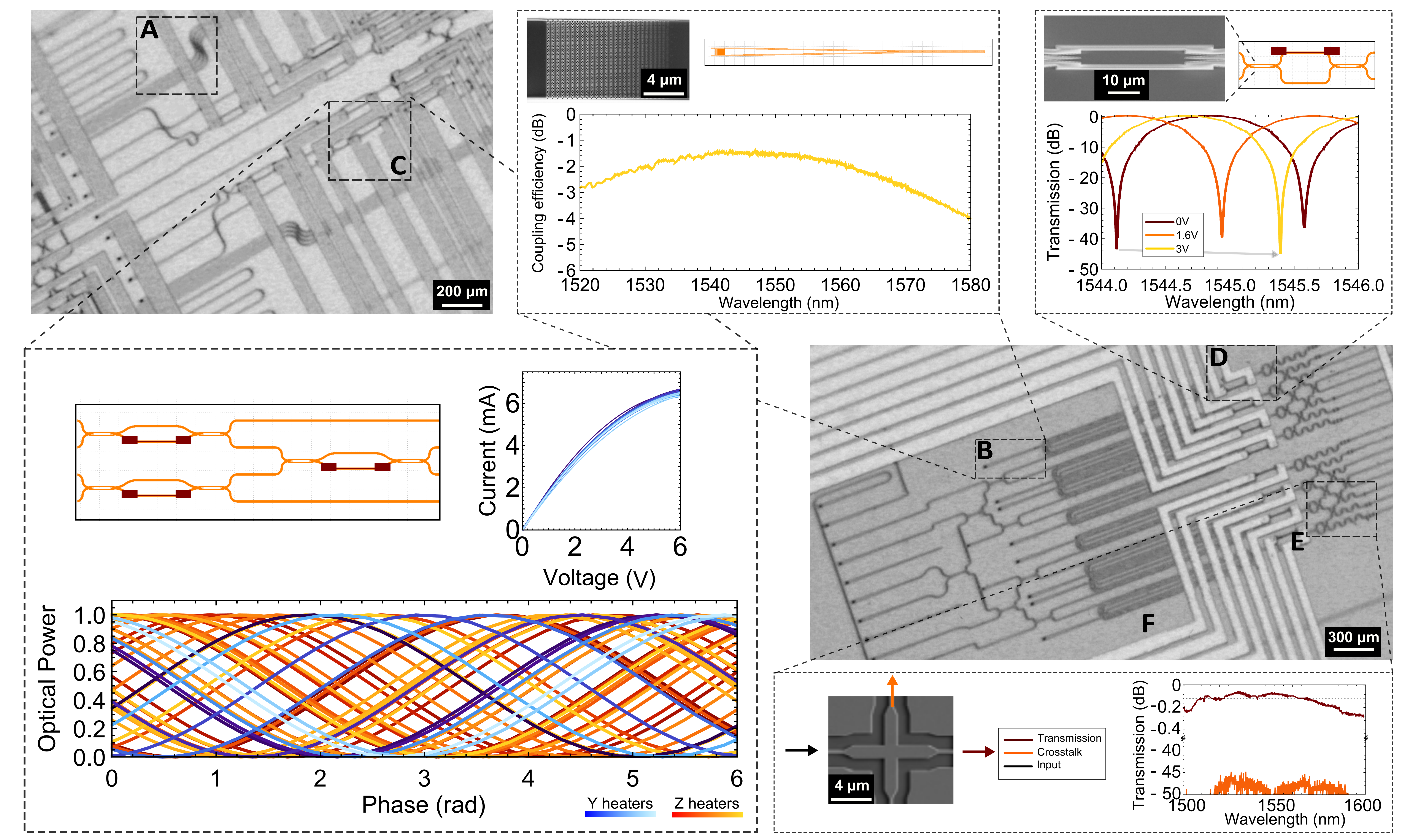}
 \caption{\textbf{Chip Components.}
 \textbf{a.} Waveguide bends and expanded waveguides.
 \textbf{b.} Grating couplers. Inset shows a SEM micrograph of the grating with its frequency response.
 \textbf{c.} Mach-Zehnder interferometers. Inset shows the current-voltage and power-phase relationships of every phase shifter on the device.
 \textbf{d.} Asymmetric Mach-Zehnder interferometers. Inset shows a SEM image of an MMI and a plot of an AMZI's transmission as a function of the input wavelength and phase shifter voltages.
 \textbf{e.} Crosser arrays. Inset shows a SEM image of a crosser and typical transmission as a function of wavelength.  
 \textbf{f.} Oblong spiral sources are designed to minimise bending losses.
  }
\label{Fig1_ChipComp}
\end{figure*} 

\subsection{Device components} 
Our device leverages transmission-optimised integrated optics throughout.
Supplementary Figure~\ref{Fig1_ChipComp} displays both microscope images of the individual chip components and plots of their characterisation.
Our grating couplers are composed of fully-etched apodized photonic crystals, further utilising Al mirrors to minimise scattering into the substrate~\cite{ding2014Gratings}. 
The waveguides on our device are fully-etched silicon waveguides with cross section $450\times250$ nm.
Propagation loss through straight waveguides was estimated via cut-back measurements to be approximately 2 dB $\text{cm}^{-1}$. 
To reduce transmission losses, long straight  waveguides are expanded via a taper to a width of 8000 nm. 
Our grating couplers have an efficiency of around -1.0\,dB  for the three wavelengths used, with 1\,dB coupling bandwidth of 40\,nm, as shown in Supplementary Figure~\ref{Fig1_ChipComp}b.

Mach-Zehnder interferometers (MZI) are realised via $2\times2$ multi-mode interferometers (MMIs) which have $< 0.5$ dB insertion loss (see Supplementary Figure~\ref{Fig1_ChipComp}c-d).
Our phase shifters are composed of Titanium micro-heaters, which locally change the temperature, and thereby the refractive index, of the silicon waveguides. 
The phase shifters on the device are $1.5\times 100\,\upmu$m (see Supplementary Figure~\ref{Fig1_ChipComp}c) and are able to access a $2\pi$ phase shift within the range $[0,6]$ V. 
The average resistance of the phase shifters on the device is $500~\Omega$.  
Current-voltage and power-phase relationships of the phase shifters are shown in Supplementary Figure~\ref{Fig1_ChipComp}c. 
Similarly to previous works~\cite{Santagati2018, Paesani2017, 16DPapero}, our heaters display non-Ohmic current-voltage characteristics, which is taken into account in their calibration (see Supplementary Section \ref{SectionAppendixDevicePerformance}).

We use asymmetric MZIs (AMZIs) with arm-length difference $\Delta L$ to implement finely tunable sinusoidal wavelength demultiplexers.
We implement $\Delta L =\lambda^2 /(n_g \text{FSR}) = 22.7 \:\upmu$m to route our chosen signal and idler frequencies to different output ports of the AMZIs (frequency difference $\Delta\lambda \simeq 10$ nm).
Supplementary Figure~\ref{Fig1_ChipComp}d shows the integrated filters' typical extinction ratio of around $-40$~dB, as well as their transmission for different applied voltages.

Along with AMZIs, arrays of crossers are needed to route signal and idler photons in their relative qudit modes.
These crossers have have loss and crosstalk values of $-0.1$ dB and -$40$ dB respectively, obtained via cut-back measurements.

The above components contribute to the overall chip schematic reported in Supplementary Figure~\ref{Fig2_FullCHipSchematics}.
\begin{figure*}[b]
  \centering
  \includegraphics[trim=0 0 0 10, width=1 \textwidth]{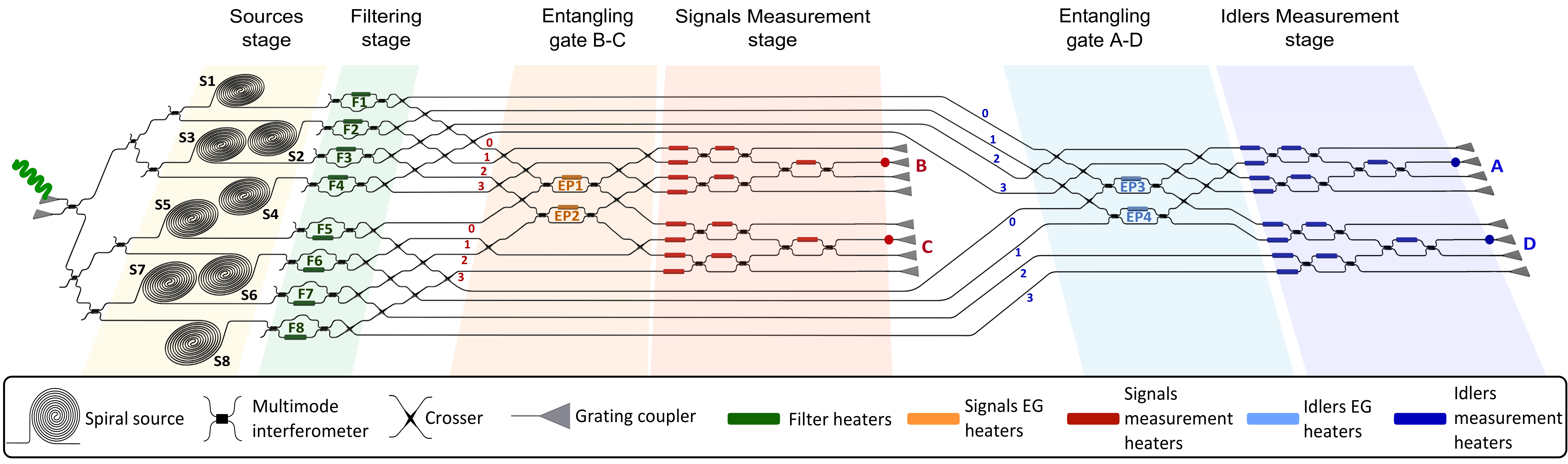}
  \caption{
  \textbf{Schematic of the device.} The chip embeds eight photon-pair sources which are coherently pumped in order to generate two pairs of non-degenerate signal and idler photons. Integrated filters at each of the sources' outputs spatially the separate signal and idler photons to obtain a biseparable four-photon state. Here, each photon pair shares four-dimensional GHZ entanglement with it's partner (A-B and C-D). Next, entangling gates A-D and B-C realise an entangled four-party four-dimensional state. Finally, a series of Mach-Zehnder interferometers enables projective measurement on each qudit.}
\label{Fig2_FullCHipSchematics}
\end{figure*} 
The overall size of the optical circuit is $\sim \, 0.2~\text{cm} \times 0.8~\text{cm}$, for a total of 48 thermo-optic titanium phase modulators and more than $200$ passive integrated components. 
The chip's maximal transmittance from the output of source S1 to output of qudit B is ${\sim} -12.9$~dB.
The average channel efficiency for a single photon from source to detector was ${\sim} -15$~dB (3$\%$), measured by the coincidences to singles ratio \cite{Silverstone2013}.
This includes the detector efficiency ($78 \%$, $-1.0$~dB), the average transmission of the off-chip filters ($87\%$, $-0.57$~dB) and fibre connection losses ($90\%$, $-0.46$~dB).\\

\subsection{Experimental set-up}

\begin{figure*}[htb]
  \centering
  \includegraphics[trim=0 0 0 10, width=1 \textwidth]{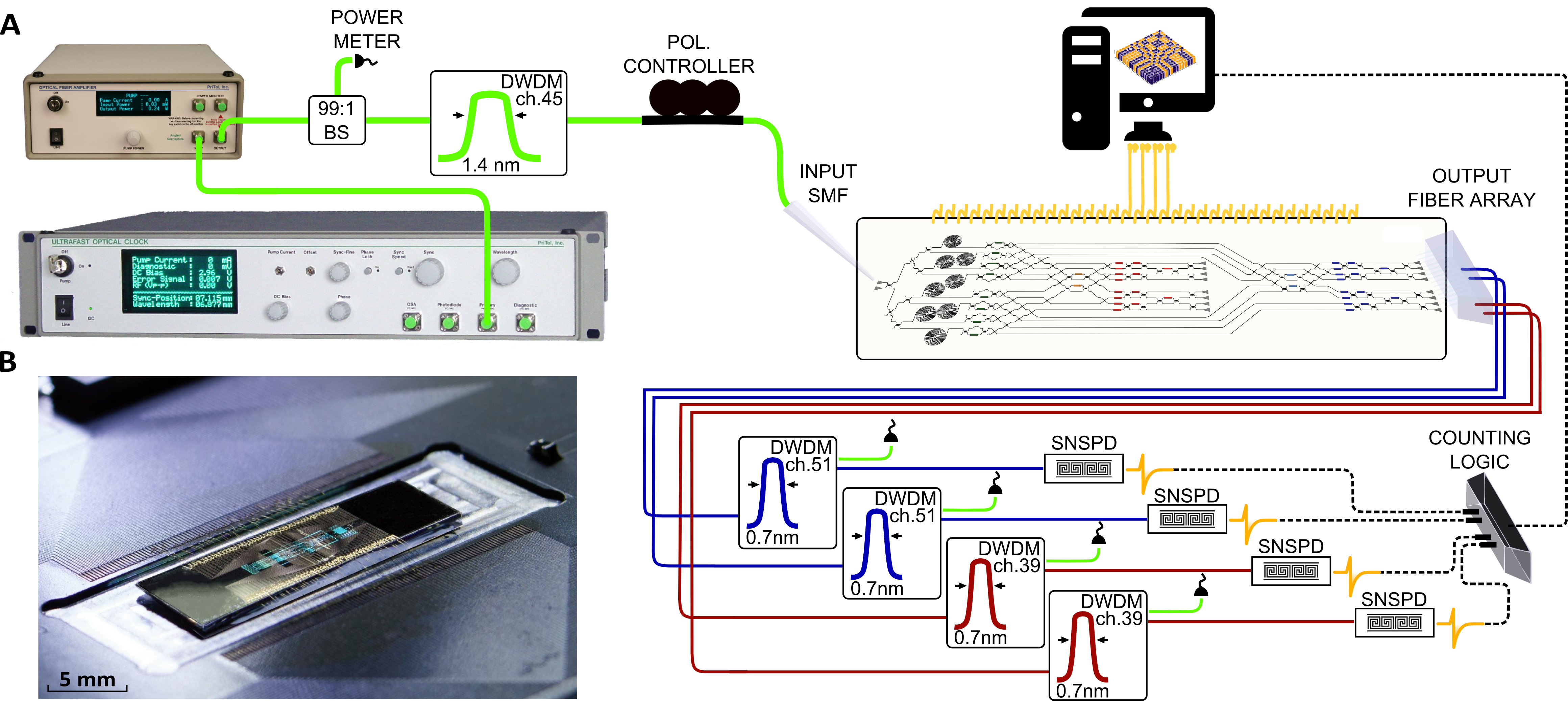}
  \caption{\textbf{Experimental setup.}
  \textbf{a.}
  Schematic showing the components and connectivity of the experimental setup.
  \textbf{b.}
  A photograph of the device.}
\label{Fig3_Suppl_ExperimentalSetup}
\end{figure*}

A schematic and photograph of the experimental setup is shown Supplementary Figure~\ref{Fig3_Suppl_ExperimentalSetup}.
Pump pulses at $1541.35$~nm ($4.80 \pm 0.03$ ps pulse duration, 500~MHz repetition rate) from an amplified erbium-doped fibre laser (Pritel) are filtered with square-shaped, $1.4$-nm-bandwidth filters (Opneti) and injected into the device with average launched pump power of $11.2$~mW. 
Signal and idler photons are collected via a fiber array and filtered with square-shaped 0.7-nm-bandwidth filters (Opneti) at $1546.12 \pm 0.35$~nm and $1536.61\pm 0.35$~nm respectively, to minimise spectral correlations and to reject the pump. 
Single photons are then detected off chip by an array of four superconducting nanowire single photon detectors with an average efficiency of $78\pm5\%$ (PhotonSpot), operating around $0.85$~K. 
Mean transmission efficiencies through the off-chip filters, averaged over the 4 channels used, are measured to be $87\%$ ($-0.57$~dB). 
Time-tags are generated (16-channel UQD-Logic) and converted to coincidences by bespoke software with a resolution of $156.25$~ps, handling count rates up to 7 MHz. 
The device is mounted using thermal epoxy and wire-bonded to an FR4 printed circuit board; temperature is stabilised using a closed-loop thermo-electric cooler (Arroyo Instruments). 
Optical coupling to fibre is achieved via a single mode fiber (SMF, input)  and via a 24-channel fibre V-groove array (output, OZ Optics). Both are controlled with 6-axis stages and piezo-electric actuators (Thorlabs). 
Analogue voltage drivers (Qontrol Systems) are used to drive the on-chip phase modulators, with 16-bit and 305~$\upmu$V resolution. 

A low input pump power (approximately $11$~mW) is used to maintain the sources' squeezing low, reducing multi-photon contamination. The typical photon pair generation probability during the measurements for each source is $|\tanh \xi|^2=0.03$, with $\xi$ the squeezing parameter. Standard four-fold coincidence rates experienced are approximately $1$\,Hz.
By adopting low loss  designs (for example directional couplers instead of multi-mode interferometers) we estimate the achievable four-fold rates to improve to the kHz regime, as described in Supplementary Section~\ref{SectionAppendixQuditEncScaling}.

\section{Device performance}\label{SectionAppendixDevicePerformance}

\subsection{Phase shifter characterisation}
The nonlinear phase-voltage relationship of the phase shifters on the device varies due to imperfections in the fabrication process. 
Therefore, each phase shifter has to be characterised individually. 
We perform this calibration in two steps. 
First, we measure the current-voltage relationship for all of the phase shifters, and fit the data to the function $I(V) = \rho_0 + \rho_1 V+\rho_2 V^2$. 
Hence $P(V) = V I = V(\rho_0 + \rho_1 V+\rho_2 V^2)$. 
Second, we enclose each phase shifter in an MZI and measure a bright-light fringe by varying the voltage applied to the phase shifter and measuring the optical power at one of the outputs.
This must be done sequentially such that each phase calibration is independent from other phase shifters on the chip.
In our calibration, these fringes have an average visibility of $99\%$.  
Examples of these fringes are displayed in Supplementary Figure~\ref{Fig1_ChipComp}c. 
The intensity of the collected light depends on the phase shifter's dissipated power through the following relation: $P_{opt}(P) = (B+A) - A \cos(\omega P - \phi_0)$, where $\phi = \omega P - \phi_0$ is the effective phase applied in the MZI.  
By fitting the fringe data with the above function we obtain $\omega$ and $\phi_0$ that allow us to reconstruct the phase-voltage relation: $\phi(V) = \omega P(V) - \phi_0 = -\phi_0 + \omega (\rho_0 V + \rho_1 V^2+\rho_2 V^3) $.  
Thermal cross-talk between the phase shifters was measured to be negligible given the relatively large size of our chip.
Finally, PID temperature control via a Peltier module prevents phase drifts due to bulk heating.\\

\subsection{Source characterisation}

\begin{figure*}[b]
  \centering
  \includegraphics[trim=0 0 0 10, width=0.93 \textwidth]{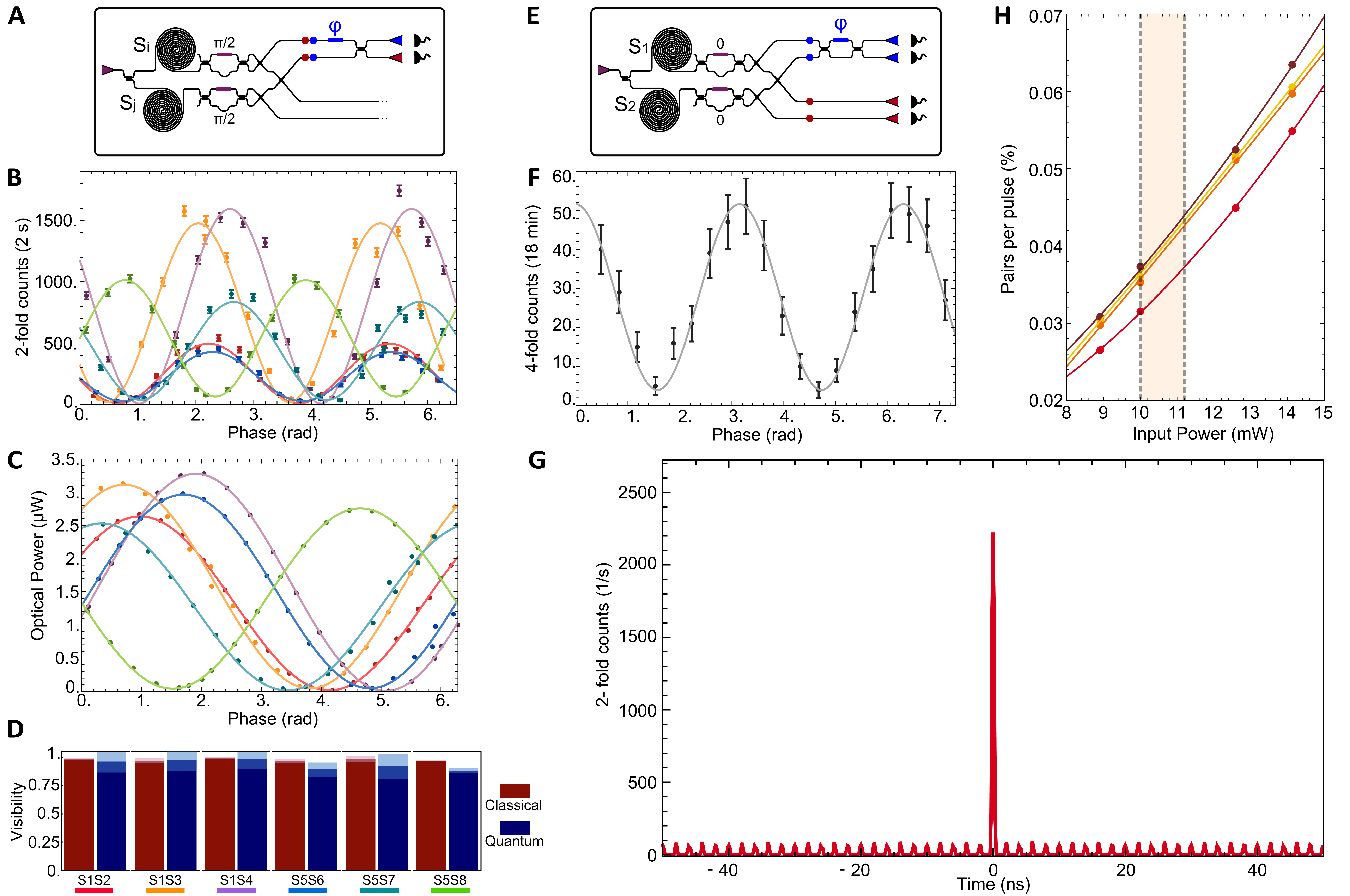}
  \caption{
  \textbf{Device characterization and sources performance.}
  \textbf{a.} Device configuration for reverse Hong-Ou-Mandel (RHOM) fringes.
  \textbf{b.} Performing single-photon detection at the output, we obtain a quantum interference fringe with a doubled frequency with respect to the classical case.
  \textbf{c.} Classical, bright-light detection pattern in the same experimental set up.
  \textbf{d.}
  RHOM interference between different choices of pairs of sources. Visibilities are all above $0.88$, showing a good degree of source indistinguishability.
  \textbf{e.} Device configuration for Heralded Hong-Ou-Mandel (HHOM) experiments.
  \textbf{f.} A HHOM fringe generated from sources S1 and S2 with a visibility of $0.86 \pm 0.04$. This corresponds to a HOM-dip equivalent visibility of $0.84\pm0.05$. The HOM-dip equivalent visibility is a lower bound on the spectral purity and total distinguishability of the photons.
  \textbf{g.} Typical histogram of the coincidence counts from one integrated source (signals and idlers from S4 and input power of ${\sim} 10$mW). 
  \textbf{h.}
  Probability of generating a photon pair for each of the sources used in GHZ-based states (sources S1, S4, S5 and S8), as a function of input laser power. The shaded region indicates the operational regime adopted in our experiments.
  All error bars are obtained assuming a Poissonian distribution of the measured counts.
  }\label{Fig4_DeviceCharacterization}
\end{figure*} 

Once the device's active components are characterised, preliminary tests on the photon-pair sources can be performed. In our chip eight 1.5-cm-long spiral silicon waveguides are present. In these, bright telecommunications-band pump pulses are converted into quantum-correlated signal and idler photons through the third-order non-linear process spontaneous four-wave mixing (SFWM).
%
Indistinguishability of photons emitted from different sources is a key characteristic to be assessed in every quantum optical implementation.  
Sources 
distinguishability 
would insert noise in the generated states, decreasing their purity, and therefore affecting their computing performances. While temporal indistinguishability is not a problem in integrated circuits, as path differences are carefully compensated during the design process, spectral indistinguishability  still represents a problem. 
For this reason off-chip narrow filters (0.7~nm bandwidth) are used and the indistinguishability is tested by performing reverse Hong-Ou-Mandel (RHOM) interference between choices of sources pairs, as shown in Supplementary Figure~\ref{Fig4_DeviceCharacterization} a-d. This represents the time reversal of the Hong-Ou-Mandel experiment: two photons in the same spatial mode (coming from the same source) impinge on a beam-splitter and come out at different ports, resulting in a coincidence peak. We can reconfigure the chip to include two chosen sources in a large MZI (Supplementary Figure~\ref{Fig4_DeviceCharacterization}a), such that by scanning the local phase associated to the mode of one of the two sources and performing single-photon coincidence detection between the two outputs, we obtain a quantum interference fringe.
We find that the resulting fringes have the signature phase-doubling associated with two-photon Fock states (Supplementary Figure~\ref{Fig4_DeviceCharacterization}b) when compared to their classical counterparts (Supplementary Figure~\ref{Fig4_DeviceCharacterization}c). By performing RHOM interference between different choices of pairs of sources we obtain a pairwise characterisation of the sources indistinguishability. 
The choices of source pairs are the same used for calibrating the phase differences among the sources: the initial $Z$ phases on modes $\ket{1}, \ket{2}, \ket{3}$ in the measurement stage of each qudit are calibrated relatively to mode $\ket{0}$, therefore with respect to the emission from source S1 and source S5. The measured visibilities of the RHOM fringes  are all above $0.88$, showing a good degree of indistinguishability (Supplementary Figure~\ref{Fig4_DeviceCharacterization}d).\\

We also perform a Heralded Hong-Ou-Mandel (HHOM) interference experiment, the chip configuration of which is shown in Supplementary Figure~\ref{Fig4_DeviceCharacterization}e.
%
Here, two signal photons from two different sources, heralded by their respective idlers, are let interfere in a MZI with variable phase $\phi$. This type of experiment informs us on both the indistinguishability of the sources (two signal photons from different sources are interfering) and on their purity (heralding of frequency correlated photons is happening at the same time, so uncorrelated joint spectral amplitude is required to have maximal fringe visibility). 
The depth of this fringe can be related to the visibility of a traditional HOM dip, and therefore the total indistinguishability of the photons, lower bounding their purity~\cite{AGGJeremy}.
A HHOM fringe obtained by recording four-fold coincidences of photons emitted by sources S1 and S2, for input power of ${\sim} 10$mW, is plotted in Supplementary Figure~\ref{Fig4_DeviceCharacterization}f. The visibility obtained from a sinusoidal fitting of the data is $0.86 \pm 0.04$, corresponding to a HOM equivalent visibility of $0.84\pm0.05$,  for a pair generation probability of ${\sim} 3\%$.  In light of these results we can confirm to have a high degree of indistinguishability and purity, comparable with state-of-the-art integrated photon-pair sources~\cite{faruque2018chip, AGGJeremy} .\\

Supplementary Figure~\ref{Fig4_DeviceCharacterization}g shows a typical histogram of coincidences at different time delays for an input power of $10$mW, which yields a pair generation probability of ${\sim} 3\%$.
Pair generation rates, for sources S1, S4, S5 and S8, as a function of the input laser power are shown in Supplementary Figure~\ref{Fig4_DeviceCharacterization}h. 
We report on these sources as they are the main ones used in graph state experiments (see Supplementary Section~\ref{SectionAppendixGraphStates}).

\section{Measurements performed for 4P4D state fidelity} \label{SectionAppendix4P4DFidelity}

Given our experimental state $\rho$, we want to calculate the state fidelity with respect to the ideal state:

\begin{equation*}
   \ket{\psi_\mathrm{4P4D}} = (\ket{0000} + e^{-i\pi/4}\ket{0033} + e^{i\pi/2}\ket{1111} + \ket{1212} + e^{i\pi/4}\ket{2121} + e^{-i\pi/4}\ket{2222} + e^{i\pi/2}\ket{3300} + e^{i\pi/4}\ket{3333})/2^{3/2}.
\end{equation*}

The fidelity can be defined as $F=Tr( \rho \ket{\psi_\mathrm{4P4D}}\bra{\psi_\mathrm{4P4D}} ) = \bra{\psi_\mathrm{4P4D}} \rho \ket{\psi_\mathrm{4P4D}} $. Only the non-zero terms of the ideal density matrix contribute to the overall fidelity (the zero terms give zero overlap with our experimental state). We can rewrite the fidelity as:
\begin{equation} F= \sum _{k l} c_l^* c_k \bra{l}\rho \ket{k} = \sum_k |c_k|^2 \bra{k}\rho \ket{k} + \sum _{k>l} [c_l^* c_k \bra{l}\rho \ket{k} + c_k^* c_l \bra{k}\rho \ket{l}] = \sum_k  |c_k|^2 \bra{k}\rho \ket{k} + \sum _{k>l} 2 Re[c_l^* c_k \bra{l}\rho \ket{k}].
\end{equation}
We need to measure, therefore, the diagonal elements ($\bra{k}\rho \ket{k}$, i.e.~the computational basis) and the off-diagonal terms $\bra{l}\rho \ket{k}$.
The projectors $\ket{l}\!\bra{k}$ are non Hermitian ($\ket{l}\!\bra{k} \neq \ket{k}\bra{l} = (\ket{l}\!\bra{k})^\dagger$), but we can rewrite them as a sum of a symmetric and an antisymmetric operators:
\begin{equation}
\ket{l}\!\bra{k} = {{\ket{l}\!\bra{k}+\ket{k}\!\bra{l}} \over {2}} + i \: {{\ket{l}\!\bra{k}-\ket{k}\!\bra{l}} \over {2i}} = \hat{M_S}+i \:\hat{M_A},
\end{equation}
where both $\hat{M_S}$ and $\hat{M_A}$ are now Hermitian.

\noindent We wish to find eigenbases for $\hat{M_S}$ and $\hat{M_A}$, in order to have a strategy for the measurement of the off-diagonal terms. Given a basis $B=\{\ket{i}\}$ with $\ket{k},\ket{l} \in B$, for any $i\neq k,l$ we have that $\hat{M_S}\ket{i}={{\ket{k}\bra{l}\ket{i}+\ket{l}\bra{k}\ket{i}}\over{2}}=0$, therefore $\ket{i}$ is an eigenvector of $\hat{M_S}$ with eigenvalue $0$. The same holds for $\hat{M_A}$: $\hat{M_A}\ket{i}=0$. On the other side, if we take $\ket{\pm_{kl}} = {{\ket{k}\pm\ket{l}}\over{\sqrt{2}}}$ and $\ket{\pm i\: _{kl}} = {{\ket{k}\pm i\: \ket{l}}\over{\sqrt{2}}}$, then $\hat{M_S}\ket{\pm _{kl}} =\pm {1\over 2} \ket{\pm _{kl}} $ and $\hat{M_A}\ket{\pm i \: _{kl}} =\pm {1\over 2} \ket{\pm i \: _{kl}} $, which makes $\ket{\pm _{kl}}$ eigenvectors of $\hat{M_S}$ with eigenvalues $\pm {1\over{2}}$ and $\ket{\pm i\: _{kl}}$ eigenvectors of $\hat{M_A}$ with eigenvalues $\pm {1\over{2}}$. The two sets $B_S = \{\ket{i}\}_{i\neq kl} \cup \{\ket{+ _{kl}}, \ket{- _{kl}}\}$ and $B_A = \{\ket{i}\}_{i\neq kl} \cup \{\ket{+ i \: _{kl}}, \ket{-i \: _{kl}}\}$ are therefore eigenbases of $\hat{M_S}$ and $\hat{M_A}$ respectively, given also their orthogonal properties.
At this point we can write:
\begin{equation}
\hat{M_S}(k,l) = {{\ket{+ _{kl}} \! \bra{+ _{kl}} - \ket{- _{kl}} \! \bra{- _{kl}}}\over{2}}
\end{equation}
\begin{equation}
 \hat{M_A}(k,l) = {{\ket{+ i \: _{kl}} \! \bra{+ i \: _{kl}} - \ket{- i \: _{kl}} \! \bra{- i \: _{kl}}}\over{2}}.
 \end{equation}
By inserting this in each of the off-diagonal terms above, we obtain:
\begin{equation}
\begin{split}
\bra{l}\rho \ket{k} &= \mathrm{Tr}(\: \rho \ket{l} \! \bra{k}) = \mathrm{Tr}(\:\rho \: M_S(k,l))+ i \:  \mathrm{Tr}(\:\rho \: M_A(k,l)) =\\
 & = {{1}\over{2}}[\mathrm{Tr}(\rho \ket{+ _{kl}} \! \bra{+ _{kl}}) - \mathrm{Tr}(\rho \ket{- _{kl}} \! \bra{- _{kl}})] +  {i\over2}[\mathrm{Tr}(\rho \ket{+ i\: _{kl}} \! \bra{+ i\: _{kl}}) - \mathrm{Tr}(\rho \ket{- i\: _{kl}} \! \bra{- i\: _{kl}})]=\\
 & =  {1\over2}[\bra{+ _{kl}} \rho \ket{+ _{kl}}- \bra{- _{kl}} \rho \ket{- _{kl}}] +  {i\over2}[\bra{+i\: _{kl}} \rho \ket{+i\: _{kl}}- \bra{-i\: _{kl}} \rho \ket{-i\: _{kl}}]=\\
 & = {1\over2}[F_{\ket{+_{kl}}}-F_{\ket{-_{kl}}}]+ {i\over2}[F_{\ket{+i\:_{kl}}}-F_{\ket{-i\:_{kl}}}],
\end{split}
\end{equation}
where $F_{\ket{+_{kl}}}$, $F_{\ket{-_{kl}}}$, $F_{\ket{+i\:_{kl}}}$ and $F_{\ket{-i\:_{kl}}}$ are the fidelities with the states $\ket{+_{kl}}$, $\ket{-_{kl}}$, $\ket{+i\:_{kl}}$ and $\ket{-i\:_{kl}}$ respectively.
Taking into account the terms of the $\ket{\psi_\mathrm{4P4D}}$ state described above, $\ket{\pm_{kl}}$ and $\ket{\pm i\:_{kl}}$ are all equivalent to GHZ states of either two or four qubits, having reduced each of the qudits to the two-mode subspace of the modes appearing in each combination of $k$ and $l$. For example the state $(\ket{2121}_{ABCD}+\ket{3300}_{ABCD})/\sqrt{2}$ arising from the term ${}_{ABCD}\bra{2121} \rho \ket{3300}_{ABCD}$ is a relabelling of a four-qubit GHZ state $(\ket{0000}_{ABCD}+\ket{1111}_{ABCD})/\sqrt{2}$ where the states $\{\ket{0},\ket{1}\}$ are represented by the modes $\{\ket{2},\ket{3}\}_{A}$, $\{\ket{1},\ket{3}\}_{B}$ , $\{\ket{2},\ket{0}\}_{C}$ and $\{\ket{1},\ket{0}\}_{D}$ for qudits A, B, C and D respectively. On the other side, combinations like $(\ket{0000}_{ABCD}+\ket{3300}_{ABCD})/\sqrt{2}$, arising from the term ${}_{ABCD}\bra{0000} \rho \ket{3300}_{ABCD}$, are equivalent to a two-qubit GHZ state multiplied by a  two-qubit product state:  $(\ket{0000}_{ABCD}+\ket{1100}_{ABCD})/\sqrt{2}$ where the states $\{\ket{0},\ket{1}\}$ in qudits A and  B are represented by the modes $\{\ket{0},\ket{3}\}_{A}$, $\{\ket{0},\ket{3}\}_{B}$.

\noindent To calculate these GHZ state fidelities we use the stabiliser technique, i.e. we measure all the stabilisers of that GHZ state, having reduced the measurements to the effective two-mode subspace occurring in each combination and performing only the minimal number of operators needed to retrieve the entire set of stabilisers.
In the case of our specific state, we need to measure 28 combinations of $\bra{l}\rho \ket{k}$. Nine operators are needed to measure all the stabilisers of a four-qubit GHZ state (see Supplementary Table~\ref{tab:stabs4GHZ})  and 3 operators for each two-qubit GHZ state stabilisers. Four states have to be measured for each combination of $k$ and $l$ ($\ket{\pm_{kl}}$ and $\ket{\pm i\:_{kl}}$), and overall 20 four-qubit GHZ state combinations and 8 two-qubit GHZ state combinations. Each four-qubit operator can be decomposed in 16 projections, while each two-qubit operator can be decomposed in 4 projections. This leads to a total of $(9 \times 4) O \times 16 P \times 20 C + (3\times 4) O \times 4 P \times 8 C = 11904$ total projectors, where $O$ stands for number of operators, $P$ stands for number of projectors and $C$ stands for the number of $k,l$ combinations. 
Considering that the stabilisers of $\ket{+_{kl}}$, $\ket{-_{kl}}$, $\ket{+ i\:_{kl}}$ and $\ket{- i\:_{kl}}$ are equivalent to the stabilisers of the star graph state up to a single qubit rotation compiled on the star central qubit, we can reduce the number of measured projectors to $ 17 O \times 16 P \times 20 C + 5 O \times 4 P \times 8 C = 5600 $. 

\noindent The overlap with the diagonal elements of the density matrix is a simple computational basis measurement (256 projectors). This is also used to normalise the off-diagonal measurements.\\

The number of measured projections is significantly lower than required for compressed sensing tomography. However, it has to be noted that this is not a full state reconstruction method, in which only the non-zero terms of the density matrix are derived in order to estimate a fidelity. Overall, only 312  of the $256 \times 256 = 65536$ density matrix elements are estimated.

\begin{table}[htb]
\centering
\begin{tabular}{	P{2cm}		P{7cm}}
Operator  &  Derived Stabilisers  \\
\hline
\hline\noalign{\vskip 1mm}    
XZZZ &  XZZZ, IIII\\ 
ZXXX &  ZXII, ZIXI, ZIIX, IXXI, IXIX, IIXX, ZXXX \\  
YYZZ &  YYZZ\\
YZYZ &  YZYZ \\
YZZY &  YZZY \\
XYYZ &  -XYYZ \\
XZYY &  -XZYY \\
XYZY &  -XYZY \\
YYYY &  -YYYY \\  
\end{tabular}
\caption{Operators measured for each of the four-qubit star stabilisers.}
\label{tab:stabs4GHZ}
\end{table}

\section{Generating and verifying graph states} \label{SectionAppendixGraphStates}

This section summarises our device's capability to generate graph and hypergraph states. 
For each state we provide an experimental recipe (source configurations and compiled qudit operations, including local complementation) and our fidelity estimation strategy. 
For each four-photon graph state we first generate the eight-qubit star and transform it using intra-qudit deterministic controlled-phase ($\hat{\mathrm{CZ}}$) gates, local complementation and $\hat{Z}$ measurements. 
We will also describe the generation and verification of the clover hypergraph state. 
Supplementary Figure~\ref{Fig12_AllClasses} lists a canonical representative graph from each of the stabiliser entanglement class this device can generate.

Using our device we can perform any qudit local operation and projective measurement. 
This provides a simple tool kit that can be used to generate a wide variety of graph states:

\begin{enumerate}
    \item Local qubit operations allow local complementation of graph states and therefore enable arbitrary transformations within entanglement classes.
    \item Qubit pairs $(1, 2)$, $(3, 4)$, $(5, 6)$ and $(7, 8)$ are encoded by the same photon. Hence, two-qubit (entangling) operations can be implemented on these qubit pairs, for example $\hat{\mathrm{CZ}}$, which toggles the presence of an edge.
    \item $\hat{Z}$ measurements, i.e.~projections onto the eigenstates of $\hat{Z}$ ($\ket{0}$ and $\ket{1}$), remove graph vertices.
\end{enumerate}

\noindent Here, local complementation is a graph operation that transforms the edges, $E$, of an input graph $G(V,E)$. Given any vertex $\alpha \in V$:
\begin{equation}
    \text{LC}_{\alpha} (G(V,E)):\rightarrow G(V,E'),
\end{equation}
such that
\begin{equation}
E' =  E \cup K_{N_G(\alpha)} - E \cap K_{N_G(\alpha)} =E \Delta K_{N_G(\alpha)}.
\end{equation}
Here $\Delta$ is the symmetric difference and $K_{N_G(\alpha)}$ are the edges of the complete graph on the neighbourhood of $\alpha$, which we write as $N_G(\alpha)$.
Informally, this ``toggles'' the edges in the subgraph of $G$ which is the neighbourhood of $\alpha$. 
That is, it adds edges which were not present, and removes them if they were present.
Local complementation can be performed on the corresponding graph state $\ket{G}$ with the application of the following local unitary:
\begin{equation}
\hat{U}^{\mathrm{LC}}_{\alpha} = \sqrt{-i\hat{X}_{\alpha}} \bigotimes_{i \in {N_G(\alpha)}} \sqrt{i \hat{Z}_{i}}.
\end{equation}
Examples of local complementation are found throughout the rest of this section.

\subsection{Two-photon, four-qubit star}
\label{sec:star2photon}

 \begin{figure*}[b]
  \centering
  \includegraphics[trim=0 0 0 10, width=1 \textwidth]{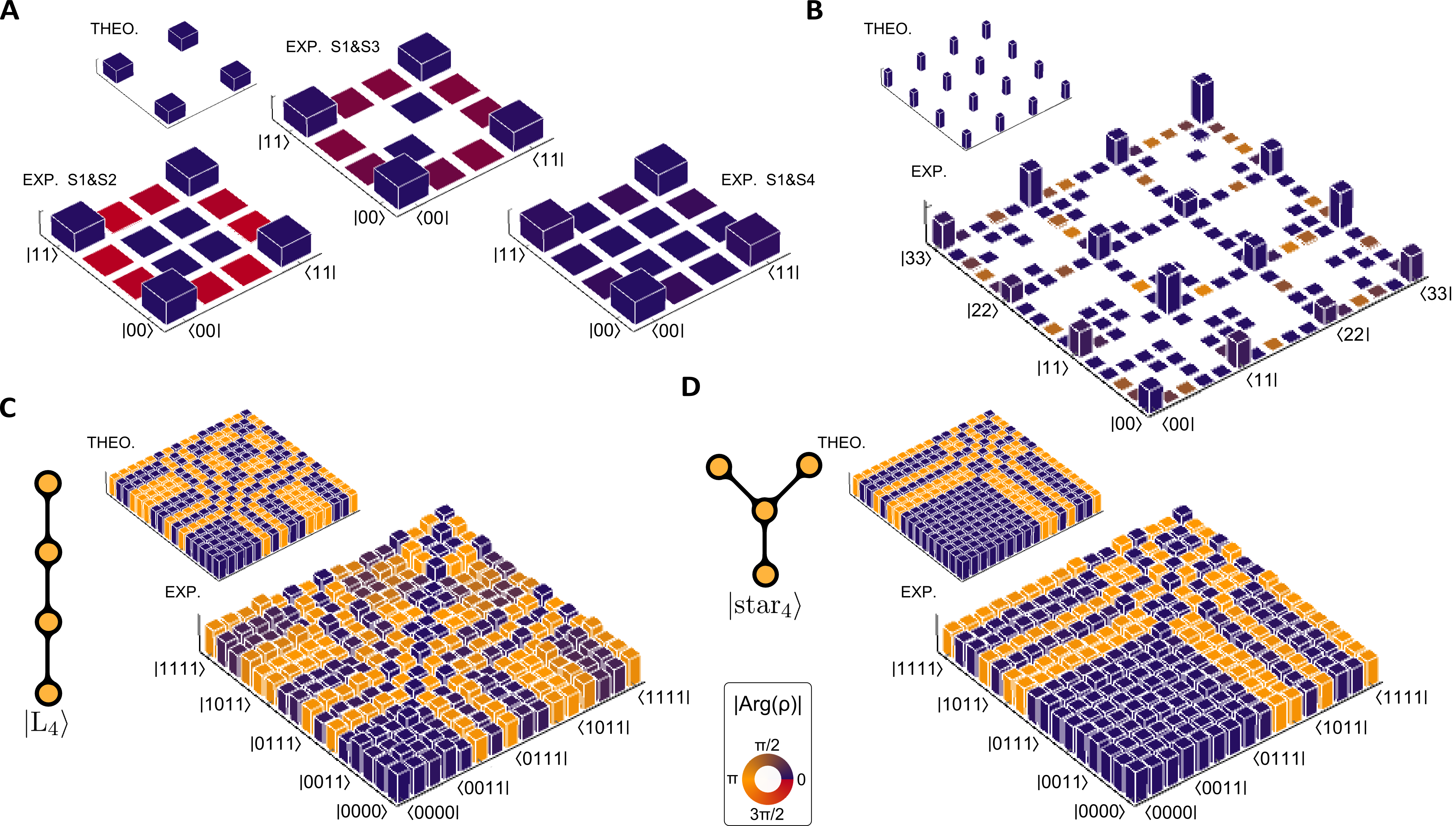}
  \caption{\textbf{Two-photon d-dimensional tomographies.}
  \textbf{a.} 
  Two-photon two-dimensional Bell-pairs generated from sources (S1, S2), (S1, S3) and (S1, S4). Tomographic reconstructions give fidelities and purities all above 0.96 and 0.94 respectively, demonstrating good indistinguishability among the sources used and good phase control among the different modes of each qudit.
  \textbf{b.} 
  Two-photon four-dimensional GHZ state generated by pumping the first four sources in a superposition. We use compressed sensing to reconstruct the experimental density matrix, which has a fidelity of $0.80\pm 0.02$ with the ideal state and a purity of $0.72\pm 0.04$.
  \textbf{c.} 
  The density matrix of our two-photon four-qubit line graph state, as reconstructed by compressed sensing tomography. This state has a fidelity $0.82\pm0.02$ with the ideal state and a purity $0.77 \pm 0.03$.
  \textbf{d.} 
  The density matrix of our two-photon four-qubit star graph state, as reconstructed by compressed sensing tomography. This state has a fidelity $0.92 \pm 0.02$ with the ideal state and a purity $0.85 \pm 0.03$.
  All error bars are obtained assuming a Poissonian distribution of the measured counts.
  }
\label{Fig5_Suppl_Tomos2PH}
\end{figure*}

To rapidly assess the functionality of the device, we perform some rudimentary two-photon state generation experiments. 
Since these only require one pair of photons to be simultaneously generated, the data collection is orders of magnitude faster than our four-photon experiments.
Moreover, we verify our ability to compile local qudit unitaries to encode four qubits in two photons (two qubits per qudit).

We generate a two-photon, four-qubit star graph state by pumping sources S1 and S4. On post-selection on having one photon per qudit we generate the following state, shown in Supplementary Figure~\ref{Fig5_Suppl_Tomos2PH}a:
\begin{equation}
    \ket{\psi} = (\ket{00}+\ket{33})/\sqrt{2}.
\end{equation}
As a state of qubits this is:
\begin{equation}
    \ket{\psi} = \ket{\mathrm{GHZ_4}} = (\ket{0000}+\ket{1111})/\sqrt{2}.
\end{equation}
Applying Hadamards $H_2H_3H_4$ yields the star state $\ket{\mathrm{star}_4}$. Our experimental $\ket{\mathrm{star}_4}$ was reconstructed using compressed sensing~\cite{gross2010CompressedSensing} (see Supplementary Figure~\ref{Fig5_Suppl_Tomos2PH}d). 
This was possible due to the high rate of data collection using two photons.

\subsection{Two-photon, four-qubit linear cluster}

The two-photon, four-qubit linear cluster state is generated by pumping sources S1 through S4. 
Post-selecting on one photon per qudit we generate the following state:
\begin{equation}
    \ket{\psi} = (\ket{00}+\ket{11}+\ket{22}+\ket{33})/2.
\end{equation} 
As a state of qubits this is:
\begin{equation}
     \ket{\psi} =  (\ket{0000}+\ket{0101}+\ket{1010}+\ket{1111})/2.
\end{equation}
We can then apply $\hat{\mathrm{CZ}}_{34}$, which is possible as qubits 3 and 4 belong to the same qudit and therefore are encoded by the same photon:
\begin{equation}
     \hat{\mathrm{CZ}}_{34}\ket{\psi} =  (\ket{0000}+\ket{0101}+\ket{1010}-\ket{1111})/2.
\end{equation}
Appyling Hadamards $H_2H_4$ yields the line graph state $\ket{\mathrm{L}_4}$, which is locally equivalent to the box graph state $\ket{\Box_4}$. 
Our experimental $\ket{\mathrm{L}_4}$ was reconstructed using compressed sensing techniques~\cite{gross2010CompressedSensing}, which were possible due to the high data collection rates associated with two-photon experiments on our device.
This reconstruction is shown in Supplementary Figure~\ref{Fig5_Suppl_Tomos2PH}c. 
The difference in the fidelity of the two-photon, four-qubit line and star graph states is explained by the different source pumping schemes used for their generation. 
The star state is obtained by pumping two sources (only one relative phase compensation), while the linear cluster state is obtained from a four-dimensional GHZ state where the overlap between four sources, and their relative calibration, contributes.
We note another method for generating the linear cluster where only two sources are used: first $\ket{\mathrm{star}_4}$ is generated (as in Supplementary Section~\ref{sec:star2photon}), then $\hat{\mathrm{CZ}}_{34}$ is applied, followed by local complementation of qubit 3.

\begin{figure}[b]
 \centering
 \includegraphics[width=1.0\textwidth]{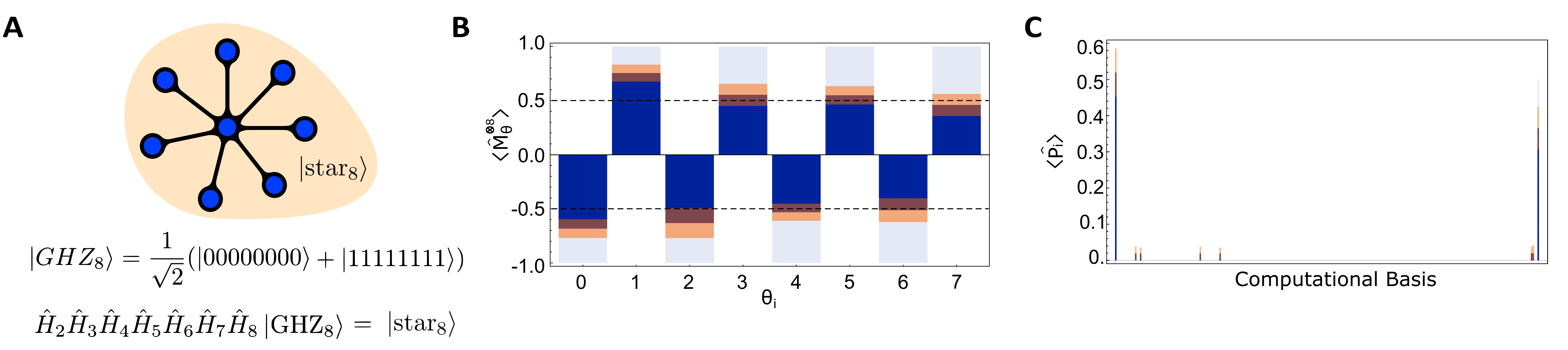}
 \caption{\textbf{The eight-qubit star graph and its verification.} \textbf{a.} The $\ket{\mathrm{star}_8}$ and its associated graph, together with its  generation recipe from the initial state $\ket{\mathrm{GHZ}_8}$.  \textbf{b.} Expectation values of the of the coherence operators $\hat{M}_\theta^{\otimes8}$. \textbf{c.} Population measurement, $P$, of $\ket{\mathrm{GHZ}_8}$. Indices are ordered as $\ket{00000000},\ket{00000001},\ldots,\ket{11111111}$.
 All error bars (shown as orange shaded regions) are obtained assuming a Poissonian distribution of the measured counts.
 }
\label{Fig6_GHZ8}
\end{figure}

\subsection{Eight-qubit star}

The eight-qubit star state is obtained by pumping sources S1, S4, S5 and S8 to generate the state:
\begin{equation}
    ((\ket{0000}+\ket{1111})/\sqrt{2})^{\otimes2} = (\ket{00000000}+\ket{00001111}+\ket{11110000}+\ket{11111111})/2,
\end{equation}
which is bi-separable. To generate a globally entangled state, we use the entangling gate B-C (see Supplementary Figure~\ref{Fig2_FullCHipSchematics}). Here, we set EP1 to perform a swap on its input modes and EP2 to perform the identity ($0$ and $\pi$ radians respectively). This removes the $\ket{00001111}$ and $\ket{11110000}$ terms by moving them outside of the post-selected, one-photon-per-qudit, basis. 
We then obtain the eight-photon GHZ state:
\begin{equation}
    \ket{\mathrm{GHZ}_8} = (\ket{00000000}+\ket{11111111})/\sqrt{2}.
\end{equation}
From here, applying Hadamards yields the eight-qubit star graph state: $\hat{H}_2\hat{H}_3\hat{H}_4\hat{H}_5\hat{H}_6\hat{H}_7\hat{H}_8\ket{\mathrm{GHZ}_8}=\ket{\mathrm{star}_8}$. Here, qubit 1 is the central qubit of the star. 
Since $\hat{Z}$ measurements can be used to remove vertices from a graph, this state can be used to produce all the star graphs of $n\leq8$ qubits.
To verify this state, we utilise the method of \cite{sackett2000GHZVerif}, that we will call $\theta$-measurements. 
First, the state is measured in the computational basis to establish the ``population", $P$. 
This is the summed probability to measure $\ket{00000000}$ or $\ket{11111111}$. 
Next, we measure the eight operators:
\begin{equation}
   \hat{M}_\theta^{\otimes8} = (\mathrm{cos}(\theta)\hat{X} + \mathrm{sin}(\theta)\hat{Y})^{\otimes8},
\end{equation}
where $\theta = k\pi/8$ for $k\in\{0, 1, 2, 3, 4, 5, 6, 7\}$. 
This corresponds to measuring each qubit in the rotated bases $\{(\ket{0}\pm e^{i \theta }\ket{1})/\sqrt{2}\}$, and is used to compute the ``coherence":
\begin{equation}
   C = \frac{1}{8}\sum_{k=0}^7 (-1)^k \expval{M_\theta^{\otimes8}}.
\end{equation}
The fidelity is then given by:
\begin{equation}
    F = \frac{1}{2}(C + P).
\end{equation}
Computational basis measurements and  $M_\theta^{\otimes8}$ expectation values are reported in Supplementary Figure~\ref{Fig6_GHZ8}, together with the state's experimental recipe.

\subsection{Linear cluster states}

The linear graph states are generated from the eight-qubit star by using the recipes shown in Supplementary Figure~\ref{Fig7_Suppl_LINEStates}.
Starting from the star, the intra-qudit $\hat{\mathrm{CZ}}_{34}$ and $\hat{\mathrm{CZ}}_{78}$ are implemented, followed by local complementations on qubits $3$ and $7$.
Then qubits $2$, $5$ and $6$ are are projected onto $\ket{0}$ to produce the five-qubit linear cluster state.
The four-qubit box cluster (which is locally equivalent to the four-qubit line) can be obtained from the five-qubit linear cluster by projecting qubit 8 onto $\ket{0}$ and subsequently performing  local complementation on qubits 3, 1 and 4. 
The three-qubit linear cluster is obtained from the five-qubit line by projecting qubits 7 and 8 onto $\ket{0}$.
\begin{figure}[h!]
 \centering
 \includegraphics[width=1 \textwidth]{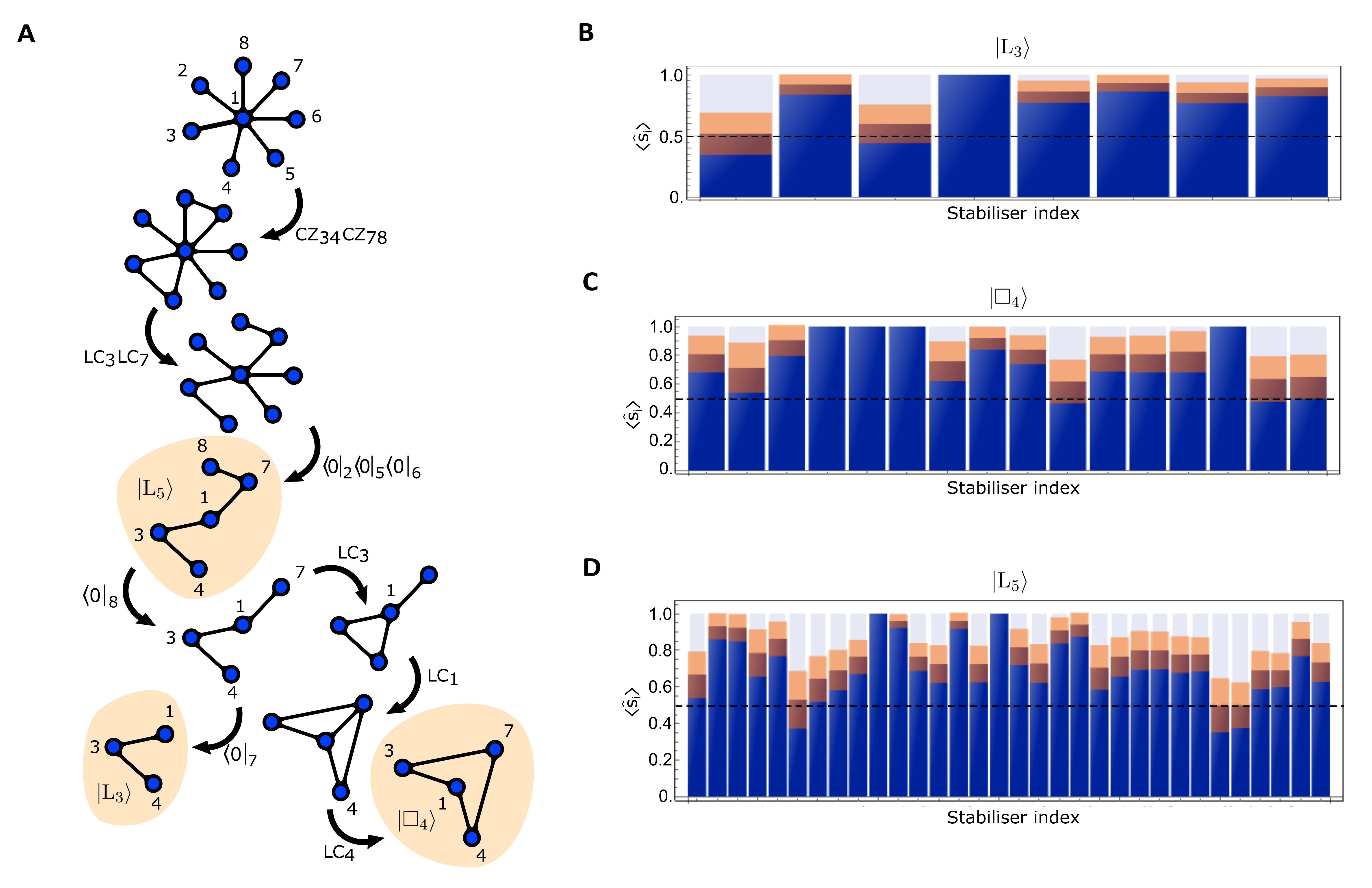}
 \caption{\textbf{Linear cluster states and their verification.} \textbf{a.} Our recipe to produce the states $\ket{\mathrm{L}_5}$, $\ket{\Box_4}$ and $\ket{\mathrm{L}_3}$. \textbf{b.} Expectation values of every stabiliser of $\ket{\mathrm{L}_3}$.  \textbf{c.} Expectation values of every stabiliser of $\ket{\Box_4}$. \textbf{d.} Expectation values of every stabiliser of $\ket{\mathrm{L}_5}$. The full list of stabiliser operators and a list of measured operators for each of the three states is reported in Supplementary Table \ref{tab:stabs5QubitLine}.
 All error bars (shown as orange shaded regions) are obtained assuming a Poissonian distribution of the measured counts.
 }
\label{Fig7_Suppl_LINEStates}
\end{figure} 

These states were verified using the operators reported in Supplementary Tables \ref{tab:stabs5QubitLine} and \ref{tab:stabs4and35QubitLine}.
Notice that the stabilisers in Supplementary Table \ref{tab:stabs5QubitLine} can be also used to contribute to a fidelity measurement on the three- and four-qubit linear graph states.
Since some stabilisers are derived from the same operator, extra measurements are required when deriving the smaller linear graphs' stabilisers.

\begin{table}[h!]
\centering
\begin{tabular}{	P{2cm}		P{9cm}}
Operator   &  Derived $\ket{\mathrm{L}_5}$ stabilisers  \\
\hline
\hline\noalign{\vskip 1mm}    
XZXZX &  IZXZI, XZIII, IIIZX, XIXZI, IZXIX, XZIZX, XIXIX, IIIII\\
ZXZXZ & ZXZII, IIZXZ, ZXIXZ \\  
YYZYY & YYZII, IIZYY, YYIYY\\
ZYYZX & ZYYIX, ZYYZI \\
XZYYZ & IZYYZ, XIYYZ \\
YXYZX & -YXYZI, -YXYIX \\
XZYXY & -IZYXY, -XIYXY \\
ZXZZX & ZXZZX \\
XZZXZ & XZZXZ \\
ZYXYZ & -ZYXYZ \\
YYIXZ & YYIXZ \\
YYZZX & YYZZX \\
XZZYY  & XZZYY \\
ZXIYY  & ZXIYY \\
ZYXXY  & ZYXXY \\
YXXYZ  & YXXYZ \\
YXXXY  & -YXXXY \\
\end{tabular}
\caption{Operators measured for all the five-qubit line stabilisers.  For ease of reading we omit hats on operators and qubit index subscripts. The Pauli operators combining each of the measured five-qubit operators are reported in qubit order (4,3,1,7,8) as per Supplementary Figure~\ref{Fig7_Suppl_LINEStates}.}
\label{tab:stabs5QubitLine}
\end{table}

\begin{table}[h!]
\centering
\begin{tabular}{	P{2cm}	P{6cm}	P{6cm}    }
Operator   &  Derived $\ket{\mathrm{L}_4}$ stabilisers &  Derived $\ket{\mathrm{L}_3}$ stabilisers \\
\hline
\hline\noalign{\vskip 1mm}    
ZXZXZ* & ZXZI, IIZX, ZXIX, IIII & - \\  
XZYYZ* & IZYY, XIYY & - \\
XZZXZ* & XZZX & - \\
ZYXYZ* & -ZYXY & - \\
YXXYZ*  & YXXY & - \\
XZXZ$^\dagger$  & IZXZ, XZII, XIXZ  &  IZX, XZI, XIX, III \\
ZYYZ$^\dagger$  & ZYYZ & ZYY \\
YYZX$^\dagger$  & YYZI, YYIX &  -  \\
YXYZ$^\dagger$  & -YXYZ & -YZY  \\
ZXZ$^\ddagger$  &  -  & ZXZ \\
YYZ$^\ddagger$  &  -  & YYZ \\
\end{tabular}
\caption{Operators measured for all the four-qubit and three-qubit line stabilisers. The Pauli operators combining each of the measured five-qubit operators are reported in qubit order (4,3,1,7,8) as per Supplementary Figure~\ref{Fig7_Suppl_LINEStates}.}
\label{tab:stabs4and35QubitLine}
\end{table}

\subsection{Double-branched and crazy graph states}%

Our three-, five- and seven-qubit branched states are generated from the eight-qubit star by using the recipe shown in Supplementary Figure~\ref{Fig8_Suppl_BranchedStates}.
First, $\hat{\mathrm{CZ}}_{34}$ is applied, then qubit 1 is local complemented. 
This produces a graph that is fully connected, but for the edge $(3,4)$.
From here, local complementing either qubit 3 or qubit 4 yields the eight-qubit branched state.
All of the other branched states can then be accessed by projecting the qubits other than 3 and 4 onto $\ket{0}$.
We choose to start with qubit 8, and go decrementally in qubit number from there.
The stabilisers generators for our double-branched graph states fidelity estimations are shown in Supplementary Table \ref{tab:stabsbranched}.
Our crazy graph state of six qubits (studied in the main text) is identical to a branched state of six qubits, and is generated in the same way. 
The recipe is shown in Supplementary Figure~\ref{Fig8_Suppl_BranchedStates}. 
The full set of stabilisers for our six-qubit crazy graph are shown in Supplementary Table \ref{tab:stabsCrazyGraph}.

\begin{figure}[h!]
 \centering
 \includegraphics[trim=0 0 0 10,clip, width=1 \textwidth]{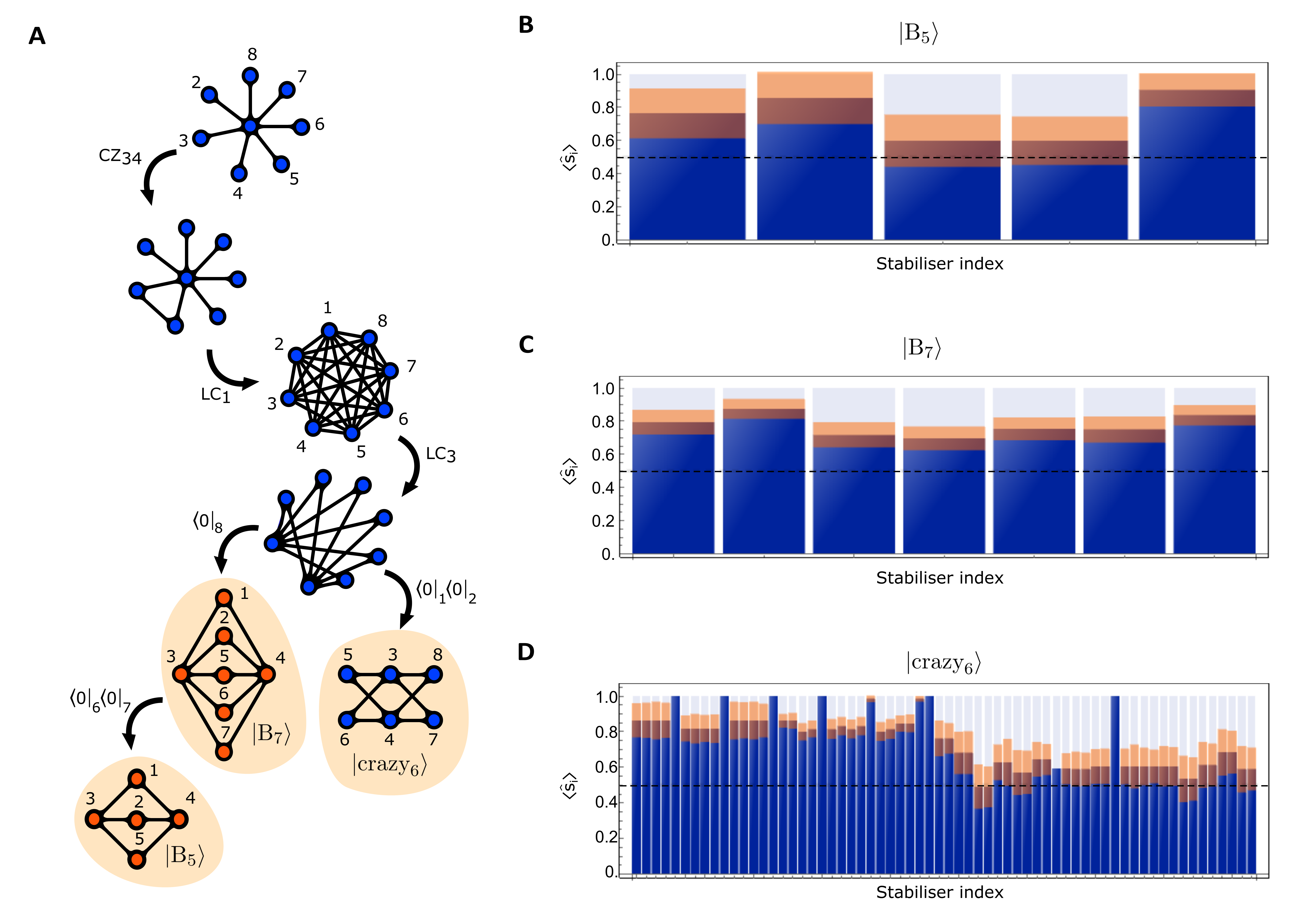}
 \caption{\textbf{Branched and crazy graph states and their verification.} \textbf{a.} Recipe to generate the states $\ket{\mathrm{B}_7}$,  $\ket{\mathrm{B}_5}$ and $\ket{\mathrm{crazy}_6}$ on our device. \textbf{b.} Expectation values of the stabiliser generators of $\ket{\mathrm{B}_5}$.  \textbf{c.} Expectation values of the stabiliser generators of $\ket{\mathrm{B}_7}$. \textbf{d.} Expectation values of every stabiliser of $\ket{\mathrm{crazy}_6}$. A full list of the measured operators are reported in Supplementary Tables  \ref{tab:stabsCrazyGraph} and \ref{tab:stabsbranched}.
 All error bars (shown as orange shaded regions) are obtained assuming a Poissonian distribution of the measured counts.
 }
\label{Fig8_Suppl_BranchedStates}
\end{figure}

\begin{table}[h!]
\centering
\begin{tabular}{P{3.5cm}	P{3.5cm}  P{3.5cm}}
 $\ket{\mathrm{B}_7}$ stabilisers  &   $\ket{\mathrm{B}_5}$ stabilisers &  $\ket{\mathrm{B}_3}$ stabilisers \\
\hline
\hline\noalign{\vskip 1mm}    
XIZZIII &  XIZZII &  XZZ  \\ 
IXZZIII &  IXZZII  &  -   \\  
ZZXIZZZ &  ZZXIZZ  & ZXI  \\
ZZIXZZZ &  ZZIXZZ &  ZIX  \\
IIZZXII &  IIZZXI &   -   \\
IIZZIXI &  IIZZIX &   -   \\
IIZZIIX &    -    &   -   \\
\end{tabular}
\caption{Branched states' stabiliser generators. The qubit order is (1, 2, 3, 4, 5, 6, 7), (1, 2, 3, 4, 5) and (1, 3, 4) for the three cases respectively, as per Supplementary Figure~\ref{Fig8_Suppl_BranchedStates}.}
\label{tab:stabsbranched}
\end{table}

\begin{table}[h!]
\centering
\begin{tabular}{	P{2cm}		P{9cm}}
Operator  &  $\ket{\mathrm{crazy}_6}$ derived stabilisers  \\
\hline
\hline\noalign{\vskip 1mm}    
XXZZXX &  XIZZII, IXZZII, IIZZXI, IIZZIX, XXIIII, XIIIXI, XIIIIX, IXIIXI, IXIIIX,  IIIIXX, XXZZXI, XXZZIX, XIZZXX, IXZZXX, XXIIXX, IIIIII\\ 
\underline{ZZ}XX\underline{ZZ} & -YYXIZZ, -YYIXZZ, -ZZXIYY, -ZZIXYY, ZXIZZ, ZZIXZZ, IIXXII, YYXIYY, YYIXYY \\  
XX\underline{XX}XX & XXXXII, XIXXXI, XIXXIX, IXXXXI, IXXXIX, IIXXXX, XXXXXX XIYYII, IXYYII, IIYYXI, IIYYIX, XXYYXI, XXYYIX, XIYYXX, IXYYXX\\
\underline{ZZ}YZZY & ZZYZZY, -YYYZZY \\
\underline{ZZ}YZYZ & ZZYZYZ, -YYYZYZ \\
\underline{ZZ}ZYZY & ZZZYZY, YYZYZY \\
\underline{ZZ}ZYYZ & ZZZYYZ, YYZYYZ \\
ZYYZ\underline{ZZ} & ZYYZZZ, ZYYZYY \\
ZYZY\underline{ZZ} & ZYZYZZ, ZYZYYY \\
YZYZ\underline{ZZ} & YZYZZZ, YZYZYY\\
YZZY\underline{ZZ} & YZZYZZ, YZZYYY\\
YZXXYZ & YZXIYZ, YZIXYZ\\
YZXXZY & YZXIZY, YZIXZY\\
ZYXXYZ & ZYXIYZ, ZYIXYZ\\
ZYXXZY & ZYXIZY, ZYIXZY\\
\end{tabular}
\caption{Measured operators for retrieving all the crazy graph stabilisers. Pairs of underlined operators correspond to intra-qudit entangled measurement settings. The qubit order is (1, 2, 3, 4, 5, 6).}
\label{tab:stabsCrazyGraph}
\end{table}


%



\clearpage
\subsection{Accessing graph state entanglement classes}
\label{FurtherGraphExplorations}

To discover which graph states our device could generate, we use a random search over the device's capabilities, starting with the star graph of 8 qubits~\cite{adcock2018postselection}.
Local complementation operations, qubit removals and controlled-Z operations between qubits (1,2), (3,4), (5,6) and (7,8) were performed at random (in software) on the star graph over many instances, until no new entanglement classes were found.
This produces a set of accessible entanglement classes, with a recipe for how to produce each one.

\begin{figure}[htb]
 \centering
 \includegraphics[trim=0 0 0 10,clip, width=0.95 \textwidth]{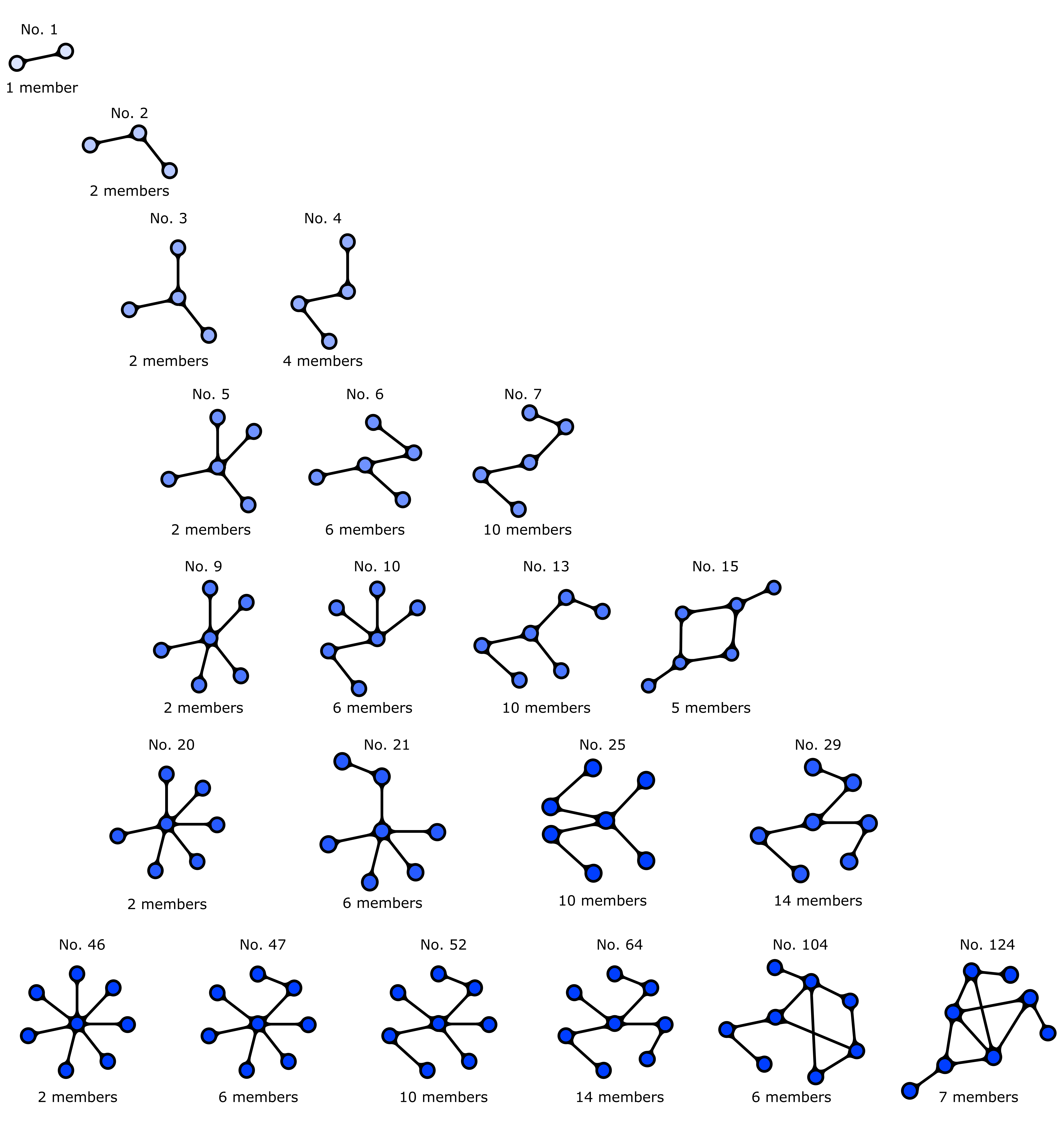}
 \caption{\textbf{Graph state classes accessible using this device.} A representative member of every graph state entanglement class accessible with this chip is shown with its index \cite{hein2004multiparty} and the number of members in that class. In total, the device can generate 21 types of stabiliser entanglement and 127 non-isomorphic graph states.}
\label{Fig12_AllClasses}
\end{figure}

\subsection{Hypergraphs}%

Our ``clover'' hypergraph state is generated by pumping sources S1 through S8 and turning on every intra-qubit gate (EP1 $\rightarrow$ 0, EP2 $\rightarrow$ 0, EP3 $\rightarrow$ 0, EP4 $\rightarrow$ 0). 
At this point the state is the same as the four-photon four-dimensional entangled state reported in the main text:
\begin{equation}
    (\ket{0000} + e^{-i\pi/4}\ket{0033} + e^{i\pi/2}\ket{1111} + \ket{1212} + e^{i\pi/4}\ket{2121} + e^{-i\pi/4}\ket{2222} + e^{i\pi/2}\ket{3300} + e^{i\pi/4}\ket{3333})/2^{3/2},
\end{equation}
the density matrix of which is reported in Figure~\ref{FigSetUp} of the main text.
Then, we apply the following unitaries to each qudit to obtain the state shown on the left in Supplementary Figure~\ref{Fig11_HypergraphRecipes}:
\begin{equation}
    \hat{U}_A = \hat{H}_1 \hat{H}_2 \hat{Z}_1 \hat{R}_{Z_2} (-\pi/4) \hat{\mathrm{CX}}_{12}
\end{equation}
\begin{equation}
    \hat{U}_B = \hat{H}_4 \hat{R}_{Z_3}(\pi/2) \hat{R}_{Z_4}(-\pi/2) \hat{\mathrm{CX}}_{34}
\end{equation}
\begin{equation}
     \hat{U}_C = \hat{Z}_5 \hat{R}_{Z_5}(\pi/4) \hat{R}_{Z_6}(\pi/2) \hat{\mathrm{CX}}_{56} \hat{\mathrm{CZ}}_{56}
\end{equation}
\begin{equation}
    \hat{U}_D =\hat{H}_7\hat{H}_8 \hat{R}_{Z_8}(-\pi/4) \hat{\mathrm{CX}}_{78}
\end{equation}
Where $\hat{H}$ is a Hadamard gate, $\hat{\mathrm{CX}}$ is a controlled-NOT two-qubit operation within the same qudit and 
\begin{equation}\hat{R}_Z(\theta) = \begin{pmatrix}
1 & 0 \\
0 & e^{i \theta}
\end{pmatrix}.
\end{equation}
This generates the eight-qubit graph state shown in Supplementary Figure~\ref{Fig11_HypergraphRecipes}, from which the clover graph state is obtained by projecting qubits $2$, $4$ and $8$ onto $\ket{0}$. 
The ``Toffoli'' and ``fully-connected Toffoli'' three-qubit hypergraphs are generated by projecting qubits $1$ and $5$ onto $\ket{00}_{15}$ and qubits $1$ and $3$ onto $\ket{11}_{13}$ respectively. 
The operators measured to acquire the full set of stabiliser expectation values for the Toffoli and fully-connected Toffoli states are reported in Supplementary Tables~\ref{tab:stabsToffoliHyperGraph} and  \ref{tab:stabsFCTOffolyHypergraph} respectively.
The two isomorphisms of the bi-separable pair of Bell states can be generated by projecting the central qubit (qubit 6) of the five-qubit clover state onto either $\ket{0}$ or $\ket{1}$, as shown in Supplementary Figure~\ref{Fig11_HypergraphRecipes}.
These two bi-separable states have stabilisers which are equivalent to each other up to swapping the measurements on the second and third qubit (see Supplementary Table \ref{tab:stabsBiseparableBellPairs}).

\begin{figure}[H]
 \centering
 \includegraphics[width=1.0 \textwidth]{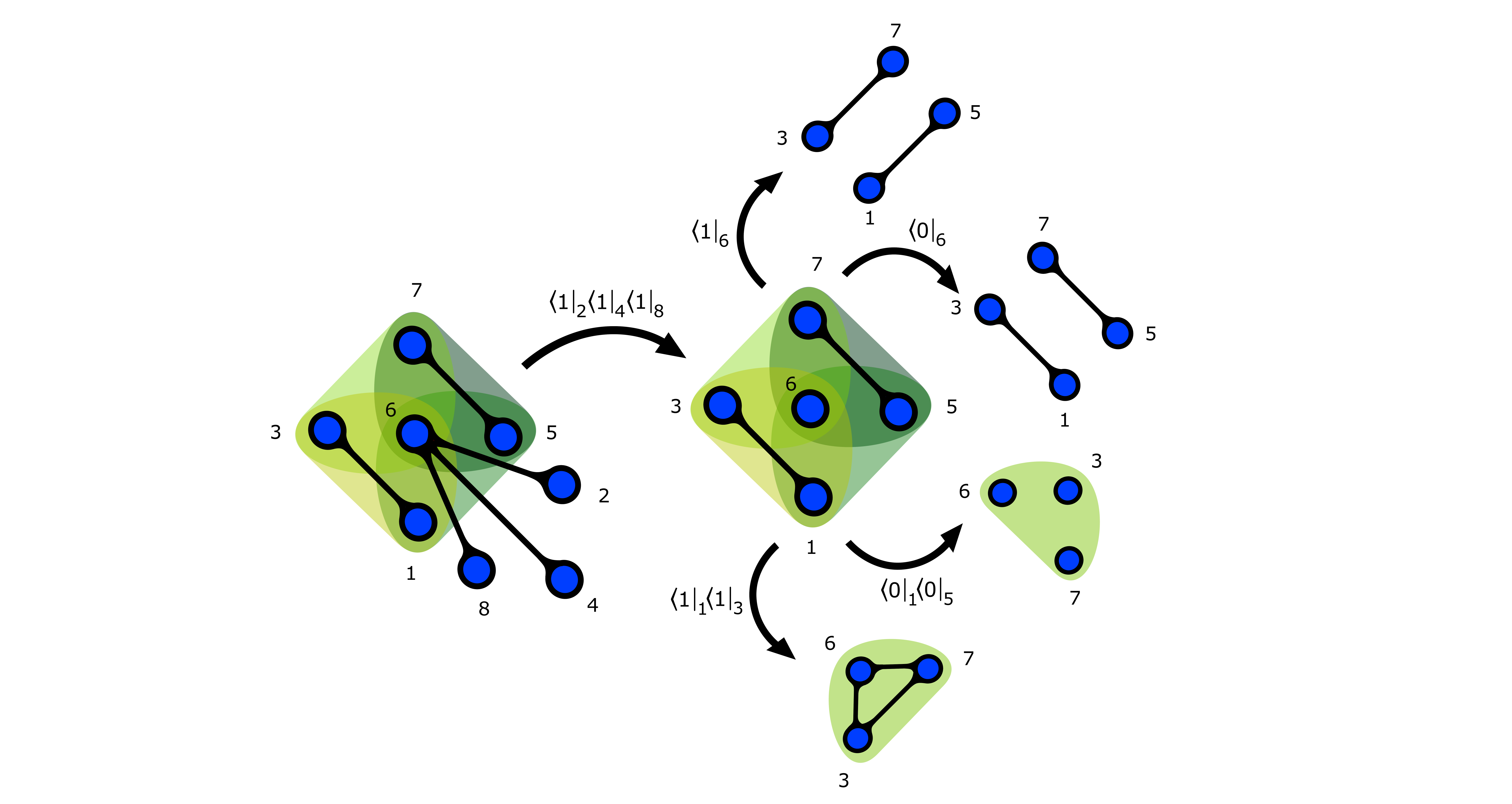}
 \caption{\textbf{Recipes for generating hypergraphs on our device.} }
\label{Fig11_HypergraphRecipes}
\end{figure}

\begin{table}[H]
\centering
\begin{tabular}{	P{4cm}		P{9cm}}
Operators  &  Toffoli hypergraph derived stabilisers  \\
\hline
\hline\noalign{\vskip 1mm}    
XZZ &  $S_1 = (\mathrm{XII}+\mathrm{XIZ}+\mathrm{XZI}-\mathrm{XZZ})/2$, $S_8 = \mathrm{I}^{\otimes 8}$\\ 
ZXZ &  $S_2 = (\mathrm{IXI}+\mathrm{IXZ}+\mathrm{ZXI}-\mathrm{ZXZ})/2$ \\  
ZZX&   $S_3 = (\mathrm{IIX}+\mathrm{IZX}+\mathrm{ZIX}-\mathrm{ZZX})/2$\\
XXZ, YYZ & $S_4 = (\mathrm{XXI}+\mathrm{XXZ}+\mathrm{YYI}-\mathrm{YYZ})/2$  \\
XZX, YZY & $S_5 = (\mathrm{XIX}+\mathrm{XZX}+\mathrm{YIY}-\mathrm{YZY})/2$  \\
ZXX, ZYY & $S_6 = (\mathrm{IXX}+\mathrm{ZXX}+\mathrm{IYY}-\mathrm{ZYY})/2$ \\
XXX, XYY, YXY, YYX & $S_7 = (\mathrm{XXX}+\mathrm{XYY}+\mathrm{YXY}+\mathrm{YYX})/2$ \\
\end{tabular}
\caption{Operators measured for the three-qubit Toffoli hypergraph state stabilisers.}
\label{tab:stabsToffoliHyperGraph}
\end{table}

\begin{table}[H]
\centering
\begin{tabular}{	P{4cm}		P{9cm}}
Operator  &  Fully-connected Toffoli hypergraph derived stabilisers  \\
\hline
\hline\noalign{\vskip 1mm}    
XZZ &  $S_1 = (-\mathrm{XII}+\mathrm{XIZ}+\mathrm{XZI}+\mathrm{XZZ})/2$, $S_8 = I^{\otimes 8}$\\ 
ZXZ &  $S_2 = (-\mathrm{IXI}+\mathrm{IXZ}+\mathrm{ZXI}+\mathrm{ZXZ})/2$ \\  
ZZX &   $S_3 = (-\mathrm{IIX}+\mathrm{IZX}+\mathrm{ZIX}+\mathrm{ZZX})/2$\\
XXZ, YYZ & $S_4 = (\mathrm{XXI}-\mathrm{XXZ}+\mathrm{YYI}+\mathrm{YYZ})/2$  \\
XZX, YZY & $S_5 = (\mathrm{XIX}-\mathrm{XZX}+\mathrm{YIY}+\mathrm{YZY})/2$ \\
ZXX, ZYY & $S_6 = (\mathrm{IXX}+\mathrm{ZXX}-\mathrm{IYY}+\mathrm{ZYY})/2$ \\
 XXX, XYY, YXY, YYX & $S_7 = (-\mathrm{XXX}-\mathrm{XYY}-\mathrm{YXY}-\mathrm{YYX})/2$ \\
         
    \end{tabular}
    \caption{Operators measured for the three-qubit fully connected Toffoli Hypergraph state stabilisers.}
    \label{tab:stabsFCTOffolyHypergraph}
\end{table}

\begin{table}[htb]
\centering
\begin{tabular}{	P{4cm}		P{4cm}}
Operator  &  Derived Stabilisers  \\
\hline
\hline\noalign{\vskip 1mm}    
XZXZ & IIXZ, XZII, XZXZ \\ 
ZXZX & IIZX, ZXII, ZXZX \\  
YYYY & YYII, IIYY, YYYY \\
XZZX & XZZX \\
ZXXZ & ZXXZ \\
YYXZ & YYXZ \\
YYZX & YYZX \\
XZYY & XZYY \\
ZXYY & ZXYY \\
         
    \end{tabular}
    \caption{Operators measured for the bi-separable state of four qubits shown in Supplementary Figure~\ref{Fig11_HypergraphRecipes}. Here, qubits $(1,2)$ and $(3,4)$ share an edge. The stabilisers for the other bi-separable state shown in Supplementary Figure~\ref{Fig11_HypergraphRecipes} can be obtained by swapping columns 2 and 3.}
    \label{tab:stabsBiseparableBellPairs}
\end{table}

\section{Measurement-based process tomography: encoding arbitrary initial states} \label{SectionAppendixEncodingMBW}

Assume we want to encode the logical state $\ket{\psi_{\mathrm{IN}}}$ in the leftmost qubit of a linear cluster state in order to apply a measurement-based quantum processes to it. Canonically, we have to prepare the first qubit of the line in the state $\ket{\psi_{\mathrm{IN}}}$ itself, entangle it with the rest of the cluster qubits with CZs, initialised in $\ket{+}$, and then measure it in the $X$ basis \cite{Raussendorf2003, vallone2008activeOWQC, Paternostro2005NoisyClusters}. We can demonstrate this is equivalent to preparing the standard cluster resource (all qubits initialised in $\ket{+}$) and measuring the first qubit in a particular basis such that the resulting initial logical state encoded is $\ket{\psi_{\mathrm{IN}}}$ itself. We show the encoding measurement procedure in the following. The initial state's basis can be written as:
\begin{equation}
\ket{\psi_{\mathrm{IN}}} = \sqrt{a} \ket{0}+\sqrt{1-a}\ e^{i\phi}\; \ket{1},
\end{equation}
\begin{equation}
\ket{\psi_{\mathrm{IN}}^{\perp}} = \sqrt{1-a}\ e^{-i\phi}\; \ket{0}-\sqrt{a}\; \ket{1}.
\end{equation}
Therefore:
\begin{equation}
\ket{0} = \sqrt{a}\; \ket{\psi_{\mathrm{IN}}}+\sqrt{1-a}\ e^{i\phi} \; \ket{\psi_{\mathrm{IN}}^{\perp}},
\end{equation}
\begin{equation}
\ket{1} = \sqrt{1-a}\ e^{-i\phi}\; \ket{\psi_{\mathrm{IN}}}-\sqrt{a}\; \ket{\psi_{\mathrm{IN}}^{\perp}}.
\end{equation}
For simplicity let us assume to have two qubits initialised in $\ket{++}_{AB}$ and entangled with a CZ operation, such that we obtain a two-qubit linear graph:
\begin{equation}
\ket{\mathrm{L}_2}_{AB} = \hat{\mathrm{CZ}}_{AB} \ket{++}_{AB} = \ket{0+}_{AB} + \ket{1-}_{AB}.
\end{equation}
We want to see the effect of an arbitrary projective measurement on the first qubit, therefore we can rewrite this as a function of the initial state's basis:
\begin{gather*}
 \ket{\mathrm{L}_2}_{AB} = (\sqrt{a}\; \ket{\psi_{\mathrm{IN}}}_A + \sqrt{1-a}\ e^{i\phi} \; \ket{\psi_{\mathrm{IN}}^{\perp}}_A )\ket{+}_{B} + (\sqrt{1-a}\ e^{-i\phi}\; \ket{\psi_{\mathrm{IN}}}_A - \sqrt{a}\; \ket{\psi_{\mathrm{IN}}^{\perp}}_A) \ket{-}_{B}  \\
 = \ket{\psi_{\mathrm{IN}}}_A (\sqrt{a}\; \ket{+}_B + \sqrt{1-a}\ e^{-i\phi} \; \ket{-}_B )+\ket{\psi_{\mathrm{IN}}^{\perp}}_A (\sqrt{1-a} \ e^{i\phi}\; \ket{+}_B - \sqrt{a} \; \ket{-}_B) \\
 = \ket{\psi_{\mathrm{IN}}}_A (\hat{H} \ket{\psi_{\mathrm{IN}}^*}_B) + \ket{\psi_{\mathrm{IN}}^{\perp}}_A (\hat{H}\ket{\psi_{\mathrm{IN}}^{\perp *}}_B).
\end{gather*}
This means that a measurement on the $\{\ket{\psi_{\mathrm{IN}}},\ket{\psi_{\mathrm{IN}}^{\perp}}\}$ basis on qubit A, results in $\{\hat{H}\ket{\psi_{\mathrm{IN}}^*},\hat{H}\ket{\psi_{\mathrm{IN}}^{\perp *}}\}$ being teleported on qubit B, therefore encoding the initial states $\hat{H}\ket{\psi_{\mathrm{IN}}^*}$ or $\hat{H}\ket{\psi_{\mathrm{IN}}^{\perp *}}$ depending on the observed output. This approach is equivalent to the standard way of encoding the logical initial state $\{\ket{\psi_{\mathrm{IN}}}\}$ of the computation in the initialisation of the leftmost qubit of the cluster. In that case we always measure in the $\hat{X}$ basis and teleport the state $\{\hat{H}\ket{\psi_{\mathrm{IN}}},\hat{H}\ket{\psi_{\mathrm{IN}}^{\perp}}\}$ (no conjugate in this case). This difference has to be taken into account when deciding which approach to take in the initialisation of the cluster and in the measurement of the initial qubit for encoding the correct logical measurement-based qubit.
As an example, we report in Supplementary Table \ref{Tab:EncodingInitialState} the required measurements for encoding the initial states needed for measurement-based process tomographies, supposing to have initialised the linear cluster in the standard way as a tensor product of $\ket{+}_i$. Generalising this to longer linear graph states is trivial.

\begin{table*}[htb]
\begin{tabular*}{0.35 \textwidth}{cc}
   Projected State &  Encoded State \vspace{1mm}\\
\hline
\hline\noalign{\vskip 1mm}    
 $\ket{0}$ & $\hat{H}\ket{0} = \ket{+}$ \\ 
 $\ket{1}$ & $\hat{H}\ket{1} = \ket{-}$ \\
 $\ket{+}$ & $\hat{H}\ket{+} = \ket{0}$ \\
 $\ket{-}$ & $\hat{H}\ket{-} = \ket{1}$ \\
 $\ket{i}$ & $\hat{H}\ket{-i} = \ket{i}$ \\
 $\ket{-i}$ & $\hat{H}\ket{i} = \ket{-i}$ \\
  \end{tabular*}
\caption{Summary of measured state projections for encoding initial states needed for measurement-based process tomographies. The initial linear cluster state is supposed to be initialised in the standard way as a tensor product of $\ket{+}_i$.}
\label{Tab:EncodingInitialState}
\end{table*}

\section{MBQC: Physical qubits to logical qubits} \label{SectionAppendixErrorCorrMBQC}

\subsection {Euler decomposition for measurement-based one-qubit rotations}

In Supplementary Table~\ref{Tab:MBAnglesLine5} we report the measurement angles to realise the reported measurement-based one-qubit gates on the five-qubit line.
\begin{table*}[htb]
\begin{tabular*}{0.7 \textwidth}{l @{\extracolsep{\fill}} cccc}
   Gate &  $\alpha$ (qubit 3) & $\beta$ (qubit 1) & $\gamma $ (qubit 7) \vspace{1mm}\\
\hline
\hline\noalign{\vskip 1mm}    
 $\hat{X}$   & $\pi \rightarrow \{\ket{-},\ket{+}\}$ & $0 \rightarrow \{\ket{+},\ket{-}\}$ & $0 \rightarrow \{\ket{+},\ket{-}\}$\\ 
 $\hat{H}$   & $\pi/2 \rightarrow \{\ket{i},\ket{-i}\}$ & $\pi/2 \rightarrow \{\ket{i},\ket{-i}\}$ & $\pi/2 \rightarrow \{\ket{i},\ket{-i}\}$\\ 
 $\hat{R}_Z(\pi/2)$   & $0 \rightarrow \{\ket{+},\ket{-}\}$ & $\pi/2 \rightarrow \{\ket{i},\ket{-i}\}$ & $0 \rightarrow \{\ket{+},\ket{-}\}$\\ 
  \end{tabular*}
\caption{Measurement angles and resulting state projections for MBQC using the five-qubit line. We report the physical qubits on which the measurements are performed in brackets, following the line generation recipe reported in Supplementary Figure~\ref{Fig7_Suppl_LINEStates}.}
\label{Tab:MBAnglesLine5}
\end{table*}

On the three-qubit line process tomography is only possible for $\hat{R}_X(\alpha)$. Supplementary Table \ref{Tab:MBAnglesLine3} reports the measurement angles and resulting state projections needed to realise the measurement-based single-qubit gates implemented via the three-qubit line reported in the main text. The mapping between physical and logical state projections for the six-qubit crazy graph is reported in Supplementary Section \ref{LogicalPauliBases}, but the same angle rules of the standard three-qubit line hold.
\begin{table*}[htb]
\begin{tabular*}{0.7 \textwidth}{l @{\extracolsep{\fill}} ccc}
   Gate &  $\alpha$ (qubit 3) & $\alpha$ (qubits 3 \& 4) \vspace{1mm}\\
\hline
\hline\noalign{\vskip 1mm}    
 $\hat{I}$ (Identity gate)   & $0 \rightarrow \{\ket{+},\ket{-}\}$ & $0 \rightarrow \{\ket{+}_\mathrm{L},\ket{-}_\mathrm{L}\}$ \\
 $\hat{X}$ (Swap gate)   & $\pi \rightarrow \{\ket{-},\ket{+}\}$ & $\pi \rightarrow \{\ket{-}_\mathrm{L},\ket{+}_\mathrm{L}\}$ \\
 $\hat{R}_X(\pi/2)$   & $\pi/2 \rightarrow \{\ket{i},\ket{-i}\}$ & $\pi/2 \rightarrow \{\ket{i}_\mathrm{L},\ket{-i}_\mathrm{L}\}$ \\
 $\hat{R}_X(-\pi/2)$   & $-\pi/2 \rightarrow \{\ket{-i},\ket{i}\}$ & $-\pi/2 \rightarrow \{\ket{-i}_\mathrm{L},\ket{i}_\mathrm{L}\}$\} \\
  \end{tabular*}
\caption{Measurement angles for the measurement-based implementation of $\hat{R}_X(\alpha)$ rotations. Measurement angles and resulting state projections for MBQC using the three-qubit line and crazy graph. Qubits 3 and (3, 4) are the central qubits of the line and crazy graph respectively (see supplementary Figures~\ref{Fig7_Suppl_LINEStates} and \ref{Fig8_Suppl_BranchedStates}).}
\label{Tab:MBAnglesLine3}
\end{table*}

\subsection {Logical Pauli bases: the crazy graph and double-branched states}
\label{LogicalPauliBases}

In the crazy graph case, at each stage of the computation the logical qubits are measured in the following logical Pauli bases:

\begin{equation}
\begin{gathered}
\ket{0}_\mathrm{L}=\ket{00}+\ket{11}=\ket{\psi^+}, \quad \ket{1}_\mathrm{L}=\ket{01}+\ket{10}=\ket{\phi^+}\\
\ket{+}_\mathrm{L}=\ket{++}, \quad \ket{-}_\mathrm{L}=\ket{--}\\
\ket{i}_\mathrm{L}=\ket{\psi^+}+i\ket{\phi^+} , \quad \ket{-i}_\mathrm{L}=\ket{\psi^+}-i\ket{\phi^+}. 
\end{gathered}
\end{equation}
The cases where one phase error occurs on the physical qubits are therefore discarded, however without any opportunity of correction (i.e.~detection only).  Note that in order to encode the $\ket{0}_\mathrm{L}$ logical state, a measurement in an entangled basis of the physical qubits is needed; this is actually possible in our case where the physical qubits are within the same qudit.

In the double-branched graphs, the middle qubits are forming one logical qubit protected under different fault scenarios. In our case we only perform the measurement-based identity, therefore we measure them in the logical $\hat{X}_\mathrm{L}$ basis:
\begin{equation}
\ket{+}_\mathrm{L}=\ket{+}^{\otimes n}, \quad \ket{-}_\mathrm{L}=\ket{-}^{\otimes n},
\end{equation}
where $n$ is the number of central qubits in the double-branched graph.

\subsection {Process tomography results}

We report every value of the corrected and uncorrected process tomographies performed on the three-qubit line in Supplementary Table \ref{Tab:OtherValuesGates}:
\begin{table*}[htb]
\begin{tabular*}{0.6 \textwidth}{l @{\extracolsep{\fill}} ccc}
   Gate &  Fidelity (physical)  & Fidelity (logical) \vspace{1mm}\\
\hline
\hline\noalign{\vskip 1mm}    
 $\hat{I}$ (Identity gate)  & $ 0.80 \pm 0.05$ & $ 0.79 \pm 0.05$ \\ 
 $\hat{X}$ (Swap gate)  & $ 0.64 \pm 0.08$ & $ 0.73 \pm 0.06$ \\ 
 $\hat{R}_X(-\pi/2)$  & $ 0.65 \pm 0.07$ & $ 0.83 \pm 0.06$  \\ 
 $\hat{R}_X(\pi/2)$ & $ 0.51 \pm 0.07$ & $ 0.77 \pm 0.09$  \\ 
  \end{tabular*}
\caption{Summary of measured fidelities of measurement-based process tomographies on physical (three-qubit line graph) and logical (six-qubit crazy graph) three-qubit line graph states.}
\label{Tab:OtherValuesGates}
\end{table*}
logical process tomography matrices are reported in the main text.

\section{Measurement-based phase estimation.}
\label{AppendixMBQCPhaseEst}
The measurement-based implementation of the phase estimation algorithm is realised by mapping the relative quantum circuit in Fig.~\href{FigMBAlgos}{\ref{FigMBAlgos}a} into measurements on the three-qubit linear graph structure shown in Fig.~\href{FigMBAlgos}{\ref{FigMBAlgos}b} as follows. Each node is measured in the Pauli X-Y plane (measurement basis $\hat{M}(\theta) = \{\ket{0} + e^{i\theta}\ket{1}, \ket{0} - e^{i\theta}\ket{1}\}$), resulting in a measurement-based gate $\hat{H}\hat{R}_Z(\theta)$. 
Therefore, measuring the leftmost and central qubits using the angles $\phi$ and $\varphi$ respectively, we obtain a measurement-based operation corresponding to $\hat{H}\hat{R}_Z(\varphi)\hat{H}\hat{R}_Z(\phi)$. Because we want the computation to be initialised in $\ket{+}$, the first measurement is performed with $\phi=0$, leading to $\hat{H}\hat{R}_Z(\varphi)\hat{H}\ket{+} = \hat{R}_X(\varphi)\ket{+}$ (also characterised in Fig.~\href{FigErrorCorrection}{\ref{FigErrorCorrection}b}). 
The rightmost measurement ($\hat{M}(\theta)$) corresponds to performing an extra ${R}_Z(\theta)$ rotation and projection in the Hadamard basis: the outcome probability measured infers the phase bit 0 or 1.
The parameters $\varphi$ and $\theta$ are iteratively chosen  depending on the outcome of the measurement on the rightmost qubit and determines the bit of the inferred phase, making the phase reconstruction process identical as the one in the circuit representation.
In the physical encoding all the measurements are performed on individual physical qubits, while in the logical encoding they correspond to non-local measurements between multiple physical qubits, as described in Supplementary Section~\ref{SectionAppendixErrorCorrMBQC}.

\section{Mapping physical errors to computational errors}
\label{SectionAppendixPhysicalErrors}

The error-correcting codes we explore in the main text do not detect arbitrary errors, but only a subsets of the possible error channels that can arise from physical noise in the device. Nevertheless, the fact that performance enhancements are experimentally observed when using logical graphs indicates that the benefits of using error-correction encodings surpass the amplification of undetected noises.
To investigate this point, in this section we numerically study how physical noise models for our silicon quantum photonic chip map into computational errors in the MBQC operations performed on the generated graph states. 
We focus on the two main sources of errors for integrated quantum photonic circuits: spectral distinguishability from the sources and imperfect control of the reconfigurable phase shifters in the interferometer.
These physical errors can in general contain both coherent and incoherent contributions and are thus difficult to map directly to bit flip or phase flip channels, or a combination of them. 
We thus build a numerical simulator so that the effects of each individual noise can be studied separately. 
To model an imperfect control of the phase-shifters, we perform a simulation of the circuit where for each phase shifter we consider the applied driving voltage obtained as an independent and identically distributed random variable  with a Gaussian distribution centered in the ideal voltage value, and with a variance $\sigma_V$ that represents the phase voltage noise. Such voltage is then mapped into that phase-shifter phase via the mapping described in Supplementary Section~\ref{SectionAppendixDevicePerformance}, using the parameters obtained from the characterisation of the phase shifters. Simulations are then repeated a large number of times (500 for the simulations reported) with independently sampled voltages, and the resulting statistics averaged to obtain fidelity estimates. The spectral distinguishability is modelled by dilating to an effective larger unitary, where partial distinguishability appears as weights for different combinations of indistinguishable interferences (see e.g. Ref.~\cite{TichyPhD}).

\begin{figure}[t!]
 \centering
 \includegraphics[width=0.9 \textwidth]{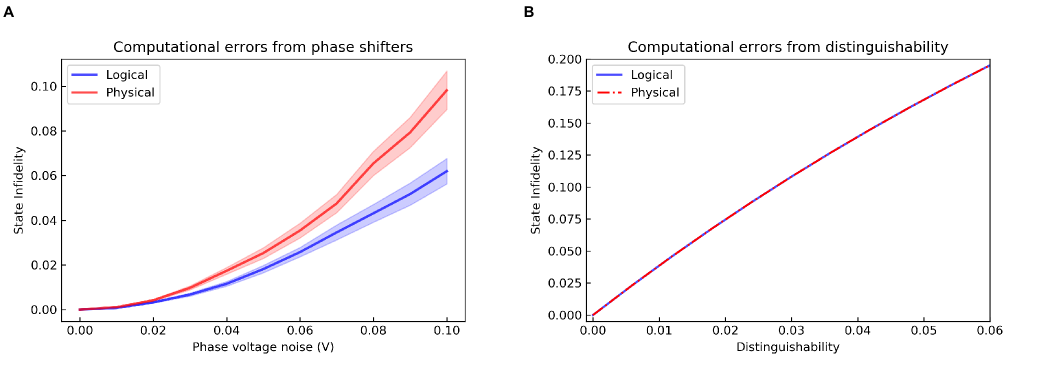}
 \caption{
 \textbf{Effects of physical noises to the infidelity of the computational states.} 
 Simulated average infidelities of the computational states obtained performing MBQC operations on physical and logical graph states as in Fig.~\ref{FigErrorCorrection}b. \textbf{a.} Average infidelities for different values of phase voltage noise for the physical (red line) and logical (blue line) encodings. Shaded areas represent $95\%$ confidence intervals from the statistics obtained from repeating the simulations with 500 in independant and identically distributed sets of noisy voltages for each data point. \textbf{b.} Average infidelities for different distinguishability values for the physical (dashed red line) and logical (blue line) encodings. For this noise model no difference between the two encodings is obtained. 
 }
\label{Fig11_Suppl_MappingPhysErrors}
\end{figure}

We investigate how these physical noises affect the MBQC operations performed on the physical and logical graph encodings reported in Fig.~\ref{FigErrorCorrection}b. We thus simulate how the average fidelities reported in Fig.~\ref{FigErrorCorrection}b are affected from the individual sources of errors. Results are reported in Fig.~\ref{Fig11_Suppl_MappingPhysErrors}, where the obtained infidelity, defined as $1-F$ where $F$ is the average state fidelity obtained, is shown for different physical noise levels for both distinguishability and phase voltage errors. Remarkably, we find that computational errors due to spectral distinguishability are independent in both physical and logical encodings (see Fig.~\ref{Fig11_Suppl_MappingPhysErrors}b). On the other hand, improvements via the logical graph encoding are observed when considering noises due to imperfect optical phase shifters, with an enhancement of approximately $4\%$ estimated for typical noise values (see Fig.~\ref{Fig11_Suppl_MappingPhysErrors}a). These results suggest that, in this experiment, the logical encoding induces a linear-optical circuit for the photonic state generation and measurements which is more resilient to imperfect linear-optical components --- in this case phase shifters.

\section{Loss tolerance in crazy graph structures}
\label{SectionAppendixSimLossTol}

One of the main challenges for photonics is reducing (or showing resilience against) photon loss. In fact, when encoding any type of large entangled quantum state, photon loss is exacerbated as the photonic circuit must be larger and more photons must be used. Moreover, since all operations are probabilistic, and the final graph state detection is usually post-selected, the overall probability of generating those states exponentially decreases as the photon number scales up.

Post-selected schemes, like the ones used in our experiments, naturally filter out photon loss. In fact, to produce the graph states reported in our work, all the four photons generated in the sources stage have to be successfully detected in their respective qudit mode to encode a meaningful qubit. The final projective measurement informs us on both the success of the graph state generation, and on the outcome of the specific measurement. Crucially, the graph state does not actually exist before all the photons are detected (and hence not lost). This comes at the expense of reducing the four-fold coincidence rate. A path towards correcting for photon loss can be found in the adoption of heralded schemes, instead of post-selected schemes, together with the same type of crazy graph encodings. In this case, conditioned by the detection of a set of heralding ancillary photons, graph states hold their existence before their actual measurement in an ``event-ready'' form, and are free to undergo any unitary or lossy process. The final projective measurement can be successful even if not all the photons in the graph are detected. Crazy graphs generated in this heralded way have been shown to be a good approach for photon loss.  Even with up to $n-1$ physical qubits lost in each $n$-qubit vertical column (corresponding to a logical qubit), there can always be found an entanglement path from the first to the last column, which allows teleportation of information \cite{RudolphOptimistic, morleyShort2019LossTol}.  These heralded approaches are also compatible with silicon photonics integrated environments and are currently still under development.  Through this work we lay the relevant groundwork for implementing these states themselves, which we expect to be heralded directly by future integrated photonics platforms that are not hindered by current experimental limitations.\\

Although working in post-selection, the loss-resilience of the graph states produced can be investigated by synthetically introducing photon loss via discarding the measured information of the associated qubits. Here, we consider the largest branched graph state created in our experiment ($\ket{B_7}$), composed of seven qubits encoded in four different photons (A, B, C and D). We perform state teleportation from the leftmost to the rightmost qubit assuming to lose one of the photons in the central error-protected layer of the graph (A, C or D), having detected the remaining three photons in a valid post-selected pattern. Here we are simulating the photon loss by partial trace: all four photons are detected to satisfy the post-selection rules, but the information about the synthetically lost photon is discarded. In Supplementary Fig.~\ref{Fig14_Suppl_SimLossTol} we show that the fidelity of the teleported state $\ket{-i}$ does not substantially change with the removal of a photon from the intermediate layer of the graph, even if the photon encodes multiple qubits. The quantum correlations remaining still allow information flow from the first leftmost qubit to the last rightmost qubit.  Interestingly, by using these crazy-graph encodings we can simultaneously show correction of phase errors and correction of photon loss. Blue and light blue points in Supplementary Fig.~\ref{Fig14_Suppl_SimLossTol} show the cases with a phase error probability $p=0.2$ and $p=0.4$ applied on all physical intermediate qubits.

\begin{figure}[t!]
 \centering
 \includegraphics[width=0.55 \textwidth]{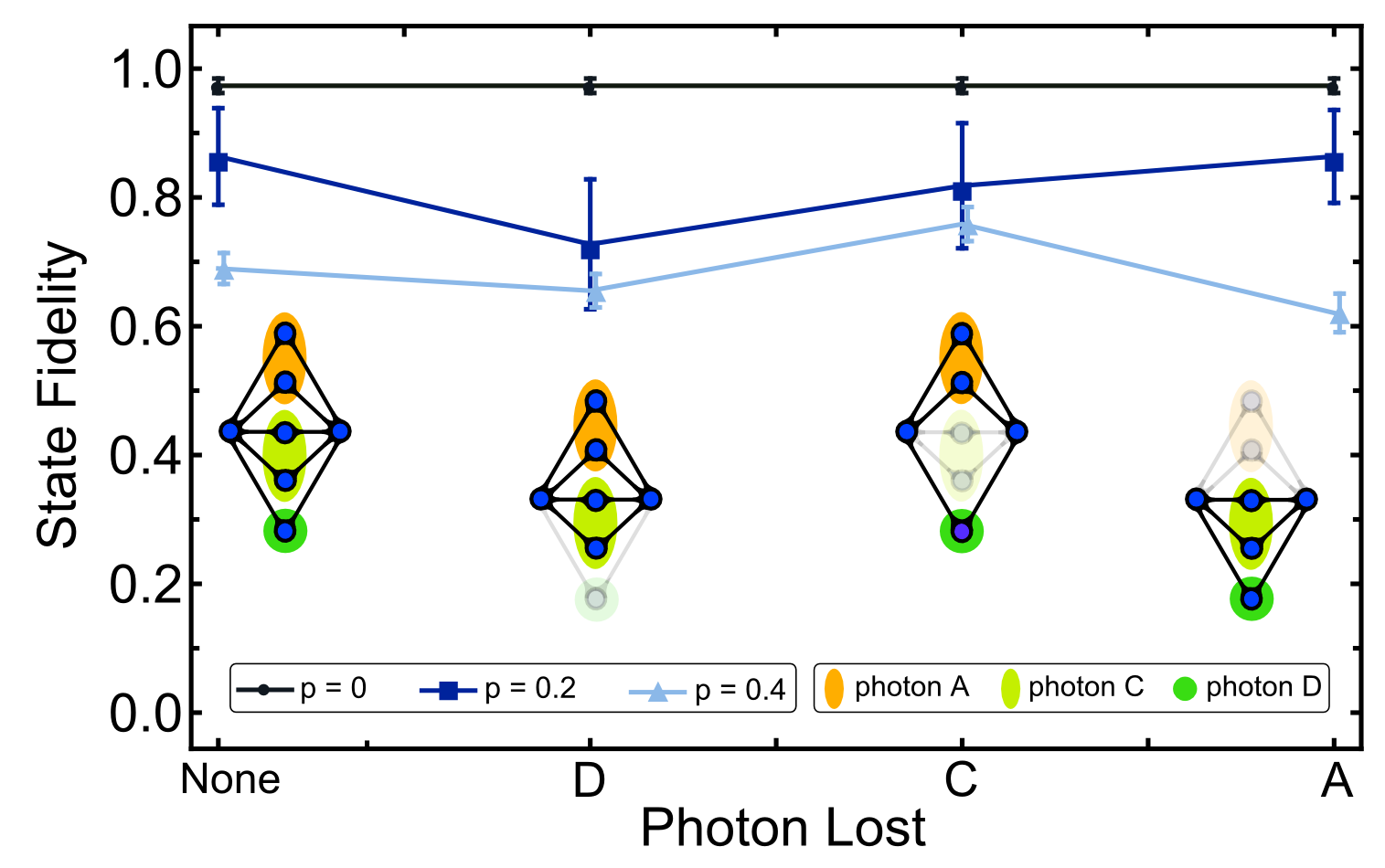}
 \caption{
 \textbf{Simulation of loss tolerance on branched states.} 
 Teleported state fidelity for $\ket{-i}$ under photon loss and continuous phase error acting on all intermediate physical qubits. Black, blue and light blue points indicate the cases with a phase error probability $p=0$, $p=0.2$ and $p=0.4$ respectively. Successful state teleportation is possible even in the case of multiple qubits being lost from one photon being lost, as it happens in our qudit encoding scheme.
 }
\label{Fig14_Suppl_SimLossTol}
\end{figure}

\section{Data analysis for comparing PEA logical and physical implementations}
\label{SectionAppendixDataAnalysis}
Here, we describe in more details the analysis performed to obtain the confidence level of achieving higher success probability when performing the phase estimation algorithm with logical qubits compared the the physical qubits implementation (see Fig.~\ref{FigMBAlgos} of the main text). We use a Monte-Carlo approach assuming binomial statistics of the photon counting events associated to each bit (i.e. the bars in Fig.~\href{\ref{FigMBAlgos}}{\ref{FigMBAlgos}c} of the main text).
The probability $p_i$ of obtaining a value 0 in the $i$-th bit is taken as the ratio between the number of measured events $n_{0,i}$ associated to the 0 outcome over the total number of events $n_i=n_{0,i}+n_{1,i}$ for that bit, with $n_{1,i}$ the number of events associated to the outcome 1. In a single round of the Monte-Carlo simulation, for each bit the number of events in 0 and 1 are resampled using a binomial distribution with number of independent events $n_i$ and success probability $p_{i}$ and $(1-p_i)$, respectively, and the phases calculated using these resampled values. The resampling is performed for both the logical and physical qubits data, and we then we calculate the number $N_\text{Log}$  and $N_\text{Phys}$ of correct phases obtained in resampled values for the logical and physical case, respectively. If $N_\text{Log}>N_\text{Phys}$, we consider the round ``won'' by the logical implementations, and by the physical implementation otherwise. After repeating this procedure for a large number $\mathcal{N}$ of rounds ($\mathcal{N}=10^4$ in our analysis), the ratio between the number of rounds won with the logical case and the total number of rounds estimates the probability that the logical implementation provides better success rates compared to the physical one in this Monte-Carlo simulation. This probability represents our estimate for the confidence level of achieving higher success probability when performing the phase estimation algorithm with logical qubits. Using the data in Fig.~\href{\ref{FigMBAlgos}}{\ref{FigMBAlgos}c} of the main text, the confidence level obtained with this analysis is $98\%$.

\section{Qudit encodings scaling for current quantum photonic implementations}
\label{SectionAppendixQuditEncScaling}

In this work we use post-selection for both photon-pair and entanglement generation, which  exponentially reduces the coincidence rate as the number of photons is increased. However, post-selection is not an essential or necessary part of linear optical quantum computing architectures: it is used here to operationally test the performance of our silicon photonic platforms when processing multi-photon entangled states, while more powerful alternatives to photon and entanglement generation are being developed (heralding and multiplexing, as discussed in the Outlook of the main text).

We here investigate what scales for NISQ-type devices are achievable in presence of post-selection. To do so, we estimate achievable rates for both our high-dimensional encoding method and standard approaches, considering inefficiencies due to post-selected probabilistic entangling gates, probabilistic photon-pair generation and losses in photonic circuitry. 
The probability of success for a post-selected entangling fusion gate, used to connect two Bell pairs of qubits or qudits for the creation of GHZ states, is $p = 1/2$. When performing multiple fusion operations in parallel, success probabilities are multiplied to calculate the total efficiency, which therefore decreases exponentially with the number of post-selected gates. In our encoding, however, the total success probability is independent on the dimension of the qudit encoded on each photon. For example, the success probability for entangling four pairs of photons to form an eight-photon GHZ graph will have a gate probability $p = (1/2)^3$.
On top of that, the probabilistic nature of photon-pair emission also has an effect on the probability of the overall state generation. Cascading many probabilistic sources (with typical photon-pair generation probabilities $s \sim 0.03$) and probabilistic gates for building larger and larger systems will inevitably imply lower probabilities of success. For example, if we start with one Bell pair, the cost of two extra photons must include an additional factor $s$: $s^2 \sim (0.03)^2$ for our spontaneous sources.
In general, when performing local photonic operations on $d$ dimensions, the photon have to pass through $d-1$ MZIs, implying that the total transmission decreases as $\eta^{d-1}$ ($\eta$ being the transmission of a single MZI).\\

%
Taking into account all the inefficiencies described above, we can calculate the generation probabilities for resource states of different sizes and compare states with the same number of qubits obtained with different schemes. In Supplementary Fig.~\ref{Fig13_Suppl_RateEstimates} we show estimates of final $n$-qubit GHZ state rates for the standard qubit encoding and the high-dimensional approach. We consider the case where components used are exactly as those in our current device (see Supplementary Section~\ref{SectionAppendixExpSetup}) as well as the case where low-loss components are used (directional couplers with $1\%$ loss, high heralding efficiency sources~\cite{Paesani2020Sources}). In both cases we consider the same laser pumping scheme used in our experiment ($s \sim 0.03$, $500$\;MHz repetition rate). For a given technology and number of photons, significant advantages are observed when adopting the high-dimensional encoding. In particular, the qudit encoding provides a feasible route towards the creation of states with up to 40 qubits, without having to push the number of processed photons beyond the current integrated multi-photon capabilities  of 10 photons.  
To achieve such scales, photon dimensionalities of up to $d=16$ are used, which have already been experimentally implemented in silicon quantum photonic~\cite{16DPapero}. Using higher dimensions should also be considered, although experimentally challenging. 
\\

\begin{figure}[t!]
 \centering
 \includegraphics[width=0.8 \textwidth]{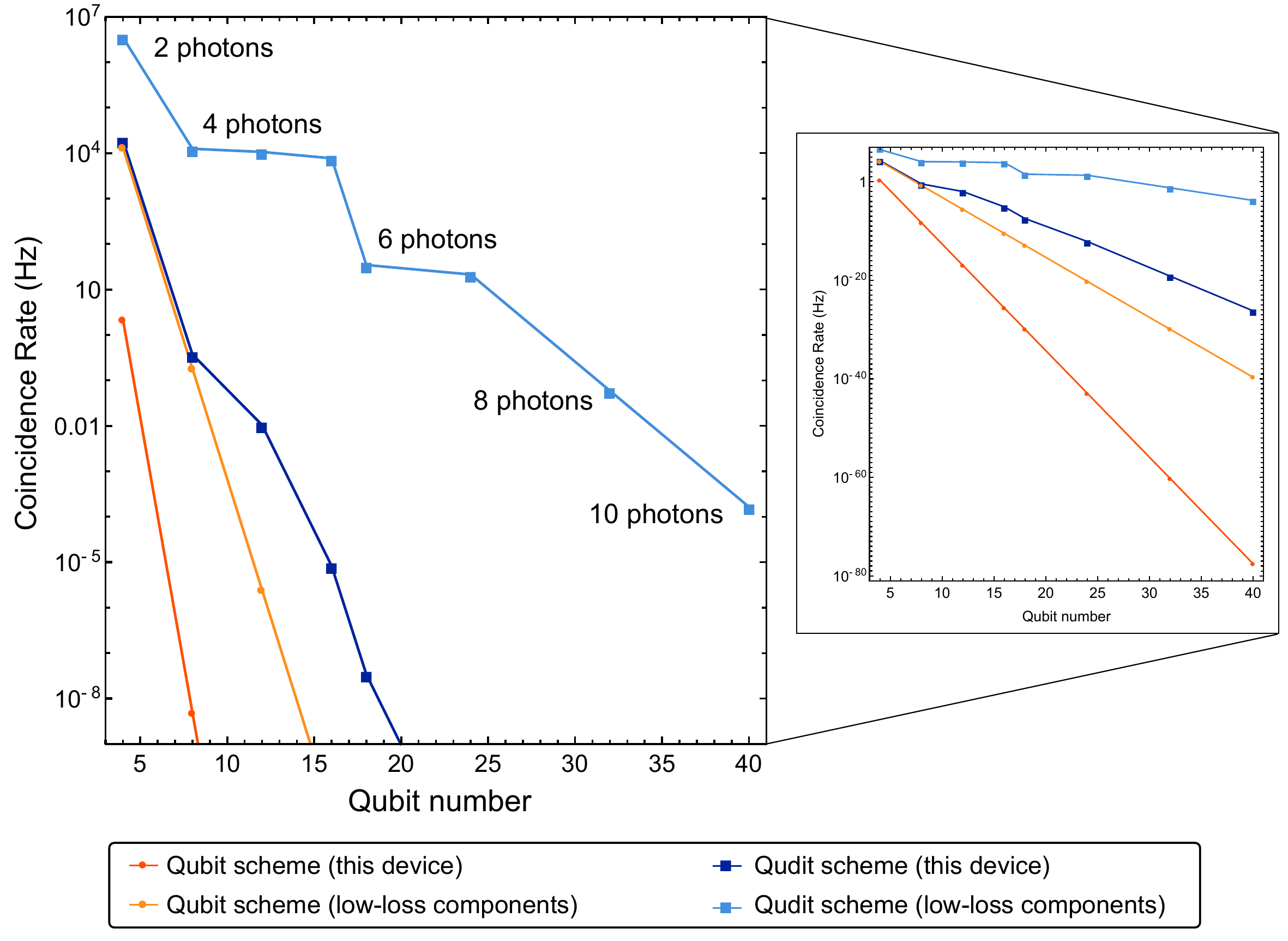}
 \caption{
 \textbf{Rate estimations for qubit and qudit encodings.} 
 Rates for $n$-qubit GHZ states obtained through qubit or qudit encodings are compared for states of different sizes. Standard expected rates are shown together with the optimal rates obtainable by assuming the best state-of-the-art chip components, chip-coupling methods and pumping schemes. When building qudit schemes equivalent to a given qubit scheme, the optimal high-dimensional scheme is chosen, having limited the number of modes to 16 and the number of photons to 10.
 }
\label{Fig13_Suppl_RateEstimates}
\end{figure}

\end{document}